\documentclass[aps,prl,showpacs,twocolumn]{revtex4-1}
\usepackage{dcolumn}
\usepackage{color}
\usepackage{amsmath,amssymb,bm, array,tabularx,booktabs,multirow}
\usepackage{longtable}

\usepackage{microtype}

\usepackage[normalem]{ulem}

\usepackage[dvipdfmx]{graphicx}
\usepackage{braket}
\usepackage{amsmath,amssymb}
\usepackage{color}

\usepackage{bm}

\usepackage{lipsum}

\renewcommand{\bibnumfmt}[1]{[#1]}
\renewcommand{\citenumfont}[1]{#1}

\begin{document}

\setlength{\extrarowheight}{4pt}


\title{High-harmonic generation in GaAs beyond the perturbative regime}

\author{Peiyu Xia$^{\ast,\dagger}$, Tomohiro Tamaya$^{\ast,\ddagger}$, Changsu Kim, Faming Lu, Teruto Kanai, Nobuhisa Ishii, Jiro Itatani, Hidefumi Akiyama, and Takeo Kato}

\thanks{$^{}$P. X. and T. T. contributed equally to this work. \\
$^{\dagger}$Correspondence: xia@issp.u-tokyo.ac.jp \\
$^{\ddagger}$Correspondence: tamaya@issp.u-tokyo.ac.jp}

\affiliation{${}$The Institute for Solid State Physics, The University of Tokyo, Kashiwa, 277-8581, Japan}

\date{\today}

\begin{abstract}

We experimentally study the field-intensity dependence of high-harmonic generation in bulk gallium arsenide in reflection geometry. 
We find the oscillatory behavior at high fields where a perturbative scaling law no longer holds. 
By constructing a theoretical framework based on the Luttinger-Kohn model, we succeed in reproducing the observed oscillatory behavior. 
The qualitative agreement between the experiment and theory indicates that field-induced dynamic band modification is crucial in the nonperturbative regime. 
We consider the origin of the oscillatory behavior in terms of dynamical localization based on the Floquet subband picture.
\end{abstract}

\maketitle

The nonperturbative properties of high-harmonic generation (HHG) in gaseous media, such as plateau and cutoff structures, originate from sub-cycle electron dynamics and can be used to produce short-wavelength attosecond pulses \cite{Corkum1993,Protopapas1997,Brabec2000,Agostini2004,Curkum2007,Krausz2009}.
Moreover, in the past decade, HHG has been experimentally observed in solids; these studies have ushered in an era of high-field condensed-matter science \cite{Ghimire2011,Schubert2014,Luu2015,Hohenleutner2015,Vampa2015,Ghimire2019}.
In contrast to gaseous media, solids have the vastly diverse nature such as in their band structures, energy gaps, crystalline anisotropy, magnetism, and so on.
HHG has presented an opportunity for generation of multi-octave coherent light that covers spectral regions from terahertz to extreme ultraviolet. 
It has thus far been investigated in various solids \cite{Ndabashimiye2016,Lanin2017,Liu2017,You2017,Yoshikawa2017,Kim2017,Jiang2018,Langer2018,VampaOE2018,Silva2018,Hirori2019,Cheng2020}, but there is as yet no theoretical framework that has enough universality to the variety of materials. 
Such a framework will be needed to transform the knowledge gained from spectroscopy of HHG in solids into novel optical technology.

It is known that the intensity of the $n$th-order harmonics obeys an $E^{2n}$ scaling law with respect to the field amplitude $E$ in perturbative nonlinear optics \cite{Yariv1984,Shen1984,Boyd2008}. 
This scaling law, however, breaks down at sufficiently high fields (typically at several MV/cm), and novel physical phenomena inherent to the nonperturbative regime emerge. 
A number of atomic experiments have observed fine-scale oscillations in the intensities of HHG as a function of the field intensity \cite{Gohle2005,Zair2008,Yost2009,Cingoz2012}.
These oscillations in HHG have been theoretically explained by quantum path interference or channel closing due to the ponderomotive shift \cite{Balcou1999,Popruzhenko2002,Kopold2002,Ishikawa2009}. 
This fact poses a question as to whether similar behavior can be observed in solids, which may illuminate the fundamental mechanism of HHG in them.

Bulk gallium arsenide (GaAs) has been intensively studied in the field of nonlinear optical physics and for applications \cite{Chin2000,Eyres2001,Mucke2001,Hirori2011,Zaks2012,Fan2013,Wismer2016,Liu2016,Schmidt2018,Schlaepfer2018,Ghalgaoui2018} because of its direct bandgap, high electron mobility, and high purity that can suppress relaxation processes. 
Recently, employment of reflection geometry has become the key to avoiding propagation effects such as phase mismatch in HHG in bulk materials \cite{VampaOE2018,Xia2018,Lu2019Nat}. 
GaAs also has an advantage in that it has a detailed theoretical framework which was developed for semiconductors \cite{Luttinger1955,Chuang1991,Ahn1995,Pfeffer1996,Pryor1998,Dargys2002,Tomic2006,Luque2015,Bastos2016,Sytnyk2018}. 
Thus, GaAs is an ideal platform for exploring HHG in solids from the perturbative to nonperturbative regime. 
For this purpose, a theory is needed that can describe the nonperturbative nature of strong-field excitation and {dynamic} band modification, which has so far been discussed only in terms of a simplified two-band model \cite{Tamaya2016,Tamaya2016PRBR,Yoshikawa2017,Tamaya2019}.

In this paper, we experimentally investigate the field-intensity dependence of HHG in GaAs. 
We observe the oscillatory behaviors in the nonperturbative regime. 
By constructing a theoretical framework based on the Luttinger-Kohn model with field-induced dynamic band modification, we succeed in reproducing this oscillatory behavior. 
This model also reproduces the crossover from the perturbative to nonperturbative regime, which reveals the underlying electronic processes.

\begin{figure}[tb]
\begin{center}
\includegraphics[width=80.0mm]{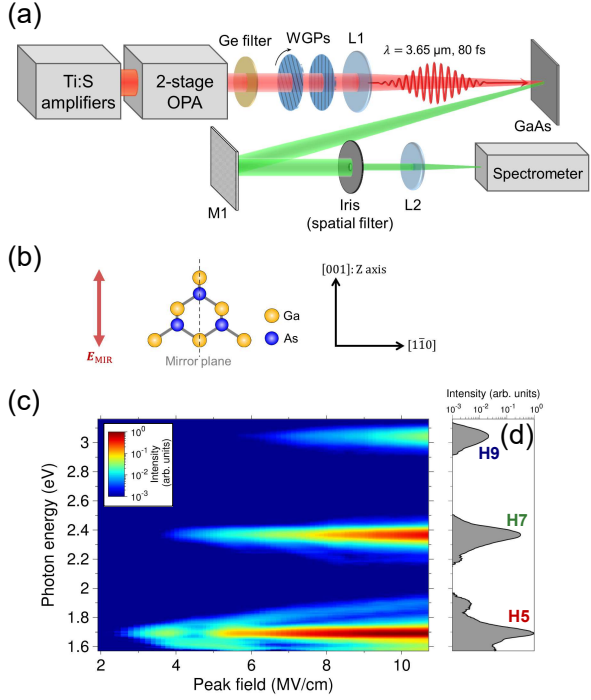}
\caption{(a) Experimental setup for HHG in reflection geometry using bulk GaAs and MIR laser source. Ti:Sapphire regenerative and multipass amplifiers (450Hz, 5 mJ, 70 fs) were used to pump the optical parametric amplifier (OPA) \cite{LuOL2018_S}. WGPs, a pair of wire-grid polarizers; L1 ($f=300$ mm) and L2 ($f=50$ mm), CaF$_2$ lenses; M1 ($R=100$ mm), Al-coated concave mirror. (b) Structure of the (110) surface of GaAs and direction of the laser polarization. (c) Field-intensity dependences of the 5th, 7th, and 9th harmonic spectra. (d) HHG spectrum at the peak field of 10 MV/cm.}
\label{fig1}
\end{center}
\end{figure}

The experiments were carried out by using an intense mid-infrared (MIR) laser irradiating a GaAs sample, as shown in Fig. \ref{fig1}(a). 
A two-stage KTiOAsO$_4$(KTA)-based optical parametric amplifier with single-plate compression \cite{LuOL2018_S} using an anitreflection-coated 5-mm-thick Ge window generated linearly polarized 80-fs pulses at 3.65 $\mu$m, which corresponds to a photon energy of 0.34 eV. 
These MIR pulses were focused on the (110) surface of a 400-$\mu$m-thick GaAs sample at room temperature and at an incidence angle of 5 degrees.
We took advantage of the reflection geometry to avoid propagation effects in the HHG process. 
A pair of wire-grid polarizers were used to adjust the laser-field intensity while keeping the linear polarization along the [001] axis. 
The peak field was estimated to be up to 12 MV/cm inside the sample, without damage. 
The HHG spectra were detected using a fiber-coupled spectrometer (QEPro, Ocean Optics), which showed only odd-order harmonics due to the inversion symmetry of GaAs (Fig. \ref{fig1}(b)).

We measured the HHG intensities integrated around each of the harmonic spectral peaks as a function of the peak field of the MIR pulses from 2 to 12 MV/cm. 
The measured field-intensity dependences were found to vary with the transverse position of the diverging HHG beam (for details, see supplementary section III), that was probably caused by the transverse intensity distribution of the MIR beam, as in an experiment in gaseous media \cite{Zair2008}. 
To avoid spatial averaging of the field-intensity dependences, we inserted an iris in the center of the high harmonic beam. 
This spatial filtering allowed us to observe the fine-scale oscillatory behavior more clearly (Fig. \ref{fig3}(a)).
It also helped to minimize the signal level of background fluorescence around the bandgap (1.42 eV) of GaAs.
In regard to the HHG spectra shown in Figs. \ref{fig1}(c) and \ref{fig1}(d), the observed harmonics often had different spectral shapes, part of which was modulated with increasing field intensity.

Figure \ref{fig3}(a) shows that the intensities of the 5th, 7th, and 9th harmonics did not saturate monotonically with increasing laser intensity, but rather exhibited oscillations. 
As the field intensity was increased, the oscillatory behaviors appeared above 4, 5, and 6 MV/cm for the 5th, 7th, and 9th harmonics, respectively, where they started to deviate from the perturbative scaling law. 
The oscillation peaks appeared well beyond the perturbative regime, and their positions varied with the harmonic order. 
Two consecutive measurements reproduced the intensities and the oscillatory behaviors, indicating no irreversible changes in the sample after each laser irradiation. 
{An explanation of these experimental results will} require clarification of the crossover of HHG from the perturbative to the nonperturbative regime as well as the physical origin of the oscillatory behavior.

\begin{figure}[t]
\begin{center}
\includegraphics[width=72.5mm]{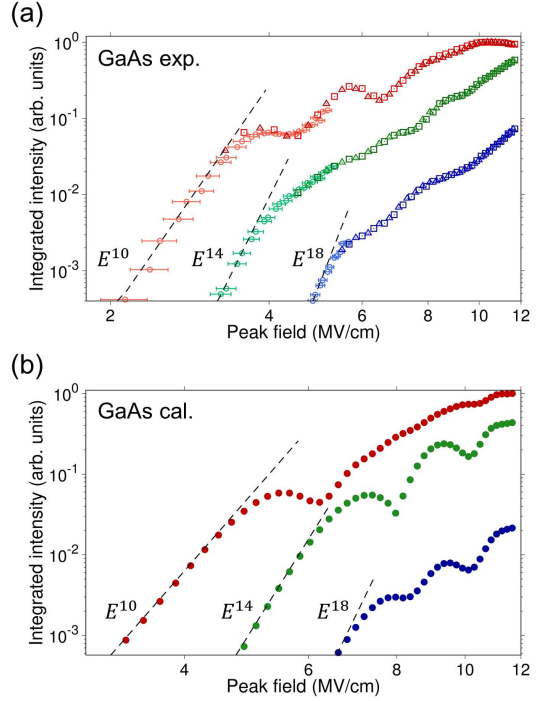}
\caption{(a) Field-intensity dependences of the 5th (red), 7th (green), and 9th (blue) HHG intensities. 
Two datasets (squares and triangles) were measured in a single sweep {starting from 2 to 12 MV/cm and then from 12 to 2 MV/cm,} respectively. 
Small HHG signals (circles) were measured with a longer acquisition time. 
(b) Calculated results of the Luttinger-Kohn model. 
In the weak-field regime, the harmonic intensities {almost} obey an $E^{2n}$ perturbative scaling law for both the measured and calculated results (dashed black lines). }
\label{fig3}
\end{center}
\end{figure}

\begin{figure}[!t]
\begin{center}
\includegraphics[width=80mm]{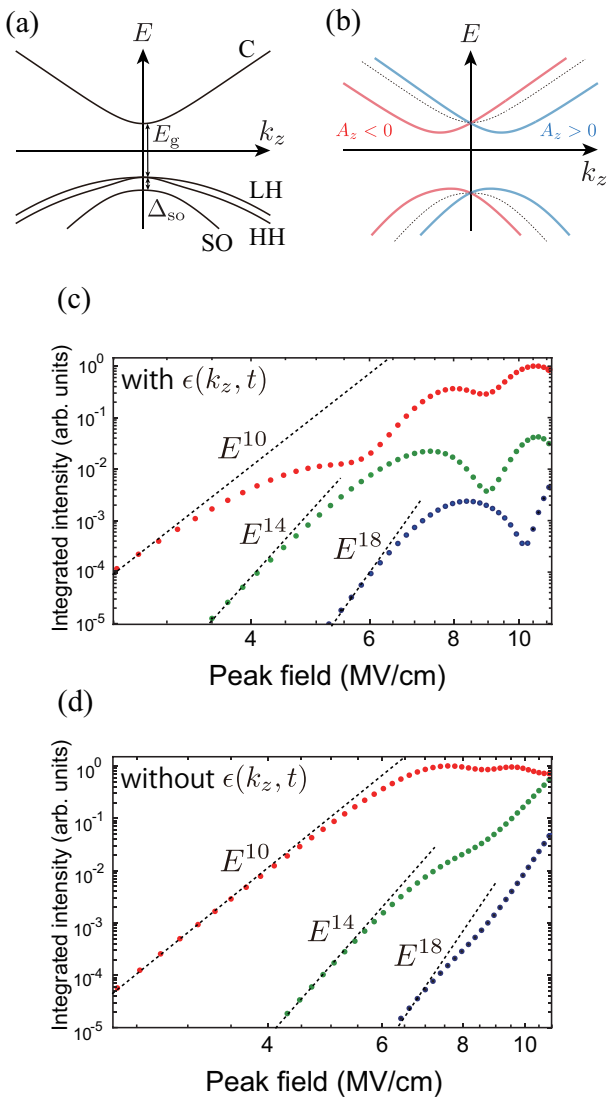}
\caption{(a) Schematic diagram of the band structure of GaAs in the absence of an external field for ${\bm k}= (0,0,k_z)$. 
The four bands, each of which is two-fold degenerate, correspond to conduction (C), light-hole (LH), heavy-hole (HH), and split-off (SO) bands. We express the bang-gap and split-off band energy as $E_\mathrm{g}$ and $\Delta_\mathrm{SO}$. 
(b) Schematic diagram of the dispersion relations of simplified model (the Kane model) under an external field. 
The blue and red lines indicate dispersion relations for $A_{z}>0$ and $A_{z}<0$. 
Numerical results of field-intensity dependence of 5th (red line), 7th (green line), and 9th (blue line) harmonics in GaAs based on the Kane model with {(c)} and without (d) {the band modification term $\epsilon(k_z,t)$}. 
These results indicate the effect of the field-induced dynamic band modification on the oscillatory behavior. }
\label{GaAs_Band_Structure}
\end{center}
\end{figure}

To analyze these experimental results, we employed an eight-band Luttinger-Kohn model, which includes conduction, heavy-hole, light-hole, and split-off bands for both spin-up and spin-down components (see Fig. \ref{GaAs_Band_Structure} (a)). 
Here, we briefly describe the model (the details are in the supplementary {material}). The model Hamiltonian is written in the form,
\begin{equation}
H_0 = \sum_{{\bm k}\sigma} \sum_{l,l'} c_{{\bm k}l \sigma'}^\dagger (H_{\bm k})_{l\sigma,l'\sigma'}
c_{{\bm k}l' \sigma'},
\end{equation}
where $c_{{\bm k}l\sigma}$ is the annihilation operator of electrons for the Bloch states constructed from the atomic states of an orbital $l(=s,p_x,p_y,p_z)$ and a spin $\sigma$. 
The matrix element $(H^{\bm k})_{l\sigma,l'\sigma'}$ is explicitly given in the supplemental material. 
We describe the external field by the vector potential ${\bm A}(t) = (0,0,A_z(t))$. The Hamiltonian for the light-matter interaction is written as
\begin{equation}
H_I = \sum_{{\bm k}\sigma} \left(-i\hbar \Omega_R(t)
c_{{\bm k}s\sigma}^\dagger 
c_{{\bm k}p_z\sigma} + {\rm h.c.}\right).
\end{equation}
Here, $\Omega_R(t) =(e/c\hbar^2)A_z(t) P_0$ is the Rabi frequency \cite{Haug2009}, and $P_0$ is the dipole matrix element \cite{Bastard1990}. 
Neglecting the dephasing effect, the wavefunction of the system can be regarded as a collection of quantum eight-state systems, which are defined at each wavenumber $\bm{k}$. 
We calculated the time evolution of the quantum state under an external field by solving the time-dependent Schr\"odinger equation {for each wavenumber} with the initial condition that all the three valence bands are occupied by electrons. 
Note that the quantum dynamics of the time-dependent Hamiltonian induces nonadiabatic excitation through the Landau-Zener transition \cite{Tamaya2016}. 
The HHG spectrum was obtained by Fourier transformation of the current induced by the external field.

Figure \ref{fig3}(b) shows the numerical results of the HHG intensities in GaAs as a function of the field intensity. 
Here, the red, green, and blue dots indicate the 5th, 7th, and 9th harmonics, respectively. 
This figure shows that the intensities of the $n$th-order harmonics follow the scaling law of perturbative nonlinear optics, $I_{n} \propto E^{2n}$, for the weak field, while they start to show oscillatory behavior in the nonperturbative regime. 
These behaviors are consistent with the experimental results (Fig. \ref{fig3}(a)), although the numerical results somewhat emphasize the dips. 
We expect that the difference may come from the incomplete spatial filtering due to the finite size of the aperture in the experiment and dephasing effects due to carrier-carrier scattering which are not considered in the present calculation.

To discuss the origin of the oscillatory behavior in Fig. \ref{fig3}, let us introduce a simplified model (the Kane model) derived from the Luttinger-Kohn model \cite{Kane1957,Bastard1990} by restricting the bands to the conduction and split-off bands (see supplementary section II). 
Neglecting the spin-flip process and abbreviating the spin index, the Hamiltonian of the Kane model reduces to a $2\times 2$ matrix of the form:
\begin{equation}
H_{\bm k}^{\rm eff} = \left( \begin{array}{cc}
E_c({\bm k}) - 2\epsilon(k_z,t) & 
-\Omega_R(t)/\sqrt{3} \\
-\Omega_R(t)/\sqrt{3} & E_v({\bm k}) 
+2\epsilon(k_z,t)/3
\end{array} \right),
\label{eq:HamiltonianKane}
\end{equation}
where $E_c({\bm k})$ and $E_v({\bm k})$ are the dispersion of the conduction and valence bands, respectively, and $\epsilon(k_z,t) = (P_0 k_z/E_g) \Omega_R(t)$ is the band modification term \cite{Tamaya2016,Tamaya2016PRBR,Yoshikawa2017,Tamaya2019}. 
Now, the diagonal element of $H_{\bm k}^{\rm eff}$ represents the band dispersion of the conduction and valence bands modified by the external field $\Omega_R(t)$ ($\propto A_z(t)$) (see Fig. \ref{GaAs_Band_Structure}(b)). 
The HHG intensities {calculated} for this model {qualitatively reproduce} the oscillatory structure, as shown in Fig. \ref{GaAs_Band_Structure}(c). 
This {result} indicates that the number of valence bands is not essential to the appearance of the oscillation. 
When we calculate the HHG intensities for an artificial Hamiltonian obtained by omitting $\epsilon({k_z},t)$, the dips are much less significant, as shown in Fig. \ref{GaAs_Band_Structure}(d). 
Therefore, the temporal change in the band dispersions represented by $\epsilon(k_z,t)$ is crucial to the appearance of the oscillatory behavior.

Let us discuss the physical origin of the oscillatory behavior. 
For a continuous wave described by $\Omega_R(t)=\Omega_{R0} \cos{\omega t}$ ($\omega$: the frequency of the incident light), the {energy} levels at each wavenumber are split into Floquet subbands \cite{Tamaya2019} {due to} the band modification term {$\epsilon(k_z,t)$}. 
Then, a matrix element for interband transition {accompanying $n$-photon absorption/emission} is renormalized, and is multiplied with $J_{{n}}(\Omega_{R0}/\omega)${, which becomes zero at specific values of $\Omega_{R0}/\omega$}. 
Therefore, we expect that the oscillatory behavior of the high harmonics reflects suppression of the effective transition probability at specific values of $\Omega_{R0}/\omega$, which is called dynamical localization or destruction of tunneling \cite{Dunlap1986,Grossmann1991,Oka2005,Lignier2007}. 
Note that the positions of the dips are influenced by various external conditions, such as the wavelength or the chirp of the incident electric field. 
This indicates that transient population dynamics caused by electronic excitation on a subcycle timescale also affects the field-intensity dependence of HHG.

In the framework of perturbative nonlinear optics, the excitation processes are described by multiphoton absorption in a fixed band structure or a virtual level, and their transition probabilities become a monotonic function of the field intensity that leads to the $I_n \propto E^{2n}$ scaling law  \cite{Yariv1984,Shen1984,Boyd2008}.
In the strong electric field, however, temporal modification of the band structures becomes significant \cite{Tamaya2016,Tamaya2016PRBR,Yoshikawa2017,Tamaya2019}, and the excitation probability is expected to show nonmonotonic behavior as a function of the field intensity. 
Actually, recent HHG experiments on ZnSe, sapphire, and Si have observed similar oscillatory behavior for the 10th (2.4 eV), 7th (11 eV), and 7th (3.9 eV) harmonics, respectively \cite{Kim2017,Lanin2017,VampaOE2018}, all of which were close to the direct-bandgap energies of the corresponding materials. 
Parts of them were fitted by a power function \cite{Kim2017} or by modeling HHG induced by the intraband current in which electrons and holes are accelerated according to Bloch's theorem \cite{Lanin2017}. 
Although these analyses were in good agreement with the observed HHG signals, the fine-scale oscillatory behavior was not reproduced. 
In contrast, our work provides clear evidence that above-bandgap HHG in GaAs exhibits oscillations in both experiment and theory. 
Therefore, our clarification of its physical origin, i.e. the field-induced dynamic band structure, will be essential to gaining a full understanding of extreme nonlinear optics in solids. 
In addition, our findings could be connected to HHG in the gas phase \cite{Gohle2005,Zair2008,Yost2009,Cingoz2012} and other high-field phenomena including dynamical localization or destruction of tunneling \cite{Dunlap1986,Grossmann1991,Oka2005,Lignier2007}, which also shows oscillatory behavior as a function of the field intensity.

In conclusion, we experimentally investigated HHG in GaAs by using reflection geometry and spatial filtering to avoid propagation effects and spatial averaging. 
We found that the intensities of the observed high harmonics did not monotonically saturate but rather exhibited oscillatory behavior with increasing field intensity. 
By constructing a theoretical framework based on the Luttinger-Kohn model to include effect of field-induced dynamic band modification, we succeeded in qualitatively reproducing this oscillatory behavior. 
By analyzing a simplified theory derived from the Luttinger-Kohn model (the Kane model), we showed that this oscillatory behavior originates from the field-induced dynamic band structures due to the diagonal elements of the time-dependent light-matter interaction matrix. 
The oscillatory behavior was related to the Floquet subband picture with the transition amplitude expressed in terms of Bessel functions \cite{Tamaya2019}, {whose} {oscillatory} behavior leads to dynamical localization. 
The findings of this paper give a basis for understanding the crossover of HHG from the perturbative to nonperturbative regime and opens up the possibility of {novel} optical technologies, such as for high-field control of higher-harmonic waves and Floquet engineering in solids.

\vspace{2mm}
The authors acknowledge support from the Japan Society for the Promotion of Science (JSPS KAKENHI Grants No. JP18H01469, JP18H05250, JP19H02623, and JP19K14624). 
H. A. also acknowledges support by the MEXT Quantum Leap Flagship Program (MEXT Q-LEAP).
P. X. was supported by the Advanced Leading Graduate Course for Photon Science (ALPS).



\newpage

\pagebreak
\widetext
\begin{center}
\textbf{\large Supplemental material for High-harmonic generation in GaAs beyond the perturbative regime}
\end{center}
\setcounter{equation}{0}
\setcounter{figure}{0}
\setcounter{table}{0}
\setcounter{page}{1}
\makeatletter
\renewcommand{\theequation}{S\arabic{equation}}
\renewcommand{\thefigure}{S\arabic{figure}}
\renewcommand{\bibnumfmt}[1]{[S#1]}
\renewcommand{\citenumfont}[1]{S#1}



\maketitle

\date{\today}

\maketitle 

\section{Theoretical analysis based on the Luttinger-Kohn model}
Here, we will present a detailed explanation of the theory employed in our main paper.
We focus on III-V zinc-blend semiconductor compounds and suppose that an AC electric field is applied along the [001] direction. 
Hereafter, we consider an eight-band system that only includes conduction, heavy-hole, light-hole, and split-off bands for spin-up and spin-down components. 
We apply conventional ${\bm{k}} \cdot {\bm{p}}$ perturbation theory to the eight-band system. 
The wavefunction of the material is described by $\Psi_{n}({\bm{k}},{\bm{x}}) = e^{i {\bm{k}} \cdot {\bm{x}}} u_{n}({\bm{k}},{\bm{x}})$, where $n$ is the band index, $\bm{k}$ is the Bloch wave vector, and $\bm{x}$ is the position.

Let us start from the microscopic Hamiltonian,
\begin{align}
H=\frac{1}{2m_0}\left(\bm{p}-\frac{e}{c} \bm{A}(t)\right)^2
+\sum_{i}V(\bm{x}-\bm{R}_{i}),
\end{align} 
where $m_{0}$ is the electron mass, $e${($<0$)} is the electron charge, $\bm{p}$ is the momentum of the bare electron, $c$ is the velocity of light, $\bm{A}(t)$ is the vector potential of the incident electric fields, and $V(\bm{x}-\bm{R}_{i})$ {is} the periodic core potential of atoms located at $\bm{R}_{i}$. 
Here, we will ignore the quasi-static energy $e^2 \bm{A}^{2}(t)/2m_{0}c^2$, which only shifts the total energy \cite{Tamaya2016PRL_S,Tamaya2016PRBR_S,Tamaya2017Science_S,Tamaya2019PRBR_S}. 
Thus, we can express the total Hamiltonian $H=H_{0} + H_{I}$, where $H_{0} = \left({\bm{p}^2}/{2 m_{0}}\right) + \sum_{i}V(\bm{x}-\bm{R}_{i})$ and $H_{I} = -\left({e}/{m_{0} c}\right) {\bm{A}(t)} \cdot {\bm{p}}$.

Next, we will derive the effective Hamiltonian $H_{\bm{k}}^{\rm {eff}}$ in ${\bm{k}} \cdot {\bm{p}}$ perturbation theory. 
Let the Hamiltonian $H_{0}$ operate on the wavefunction $\Psi_{n}({\bm{k}},{\bm{x}}) = e^{i {\bm{k}} \cdot {\bm{x}}} u_{n}({\bm{k}},{\bm{x}})$, that is,
\begin{eqnarray}
H_{0} \ket{\Psi_{n}({\bm{k}},{\bm{x}})} = e^{i {\bm{k}} \cdot {\bm{x}}} \left[\frac{{\bm {p}}^2}{2 m_{0}} + \sum_{i}V(\bm{x}-\bm{R}_{i}) + \frac{\hbar}{m_{0}}{\bm{k}} \cdot {\bm{p}} + \frac{\hbar^2 \bm{k}^2}{2 m_{0}} \right] \ket{ u_{n}({\bm{k}},{\bm{x}})}. 
\end{eqnarray}
This equation provides the single-particle part of the effective Hamiltonian $H_{0,{\bm{k}}}^{\rm{eff}}$ operating on the state space of $u_{n}({\bm{k}},{\bm{x}})$, which can be expressed as
\begin{eqnarray}
H^{\rm{eff}}_{0,\bm{k}} \equiv e^{-i {\bm{k}} \cdot {\bm{x}}} H_{0} e^{i {\bm{k}} \cdot {\bm{x}}} = \frac{{\bm {p}}^2}{2 m_{0}} + \sum_{i}V(\bm{x}-\bm{R}_{i}) + \frac{\hbar}{m_{0}}{\bm{k}} \cdot {\bm{p}} + \frac{\hbar^2 \bm{k}^2}{2 m_{0}}.
\end{eqnarray}
By applying the Hamiltonian operator $H_{I}$ to the same wavefunction, we obtain the following form:
\begin{eqnarray}
H_{I} \ket{\Psi_{n}({\bm{k}},{\bm{x}})} = -\frac{e}{m_{0}c}e^{i {\bm{k}} \cdot {\bm{x}}} \left[{\bm {A}}(t) \cdot \hbar \bm{k} + \bm{A}(t) \cdot {\bm{p}} \right] \ket{ u_{n}({\bm{k}},{\bm{x}})}.
\end{eqnarray}
This relationship suggests the following effective Hamiltonian, 
\begin{eqnarray}
H^{\rm{eff}}_{I,\bm{k}} \equiv e^{-i {\bm{k}} \cdot {\bm{x}}} H_{I} e^{i {\bm{k}} \cdot {\bm{x}}} = -\frac{e}{m_{0}c} \left[{\bm {A}}(t) \cdot \hbar \bm{k} + \bm{A}(t) \cdot {\bm{p}} \right]
\end{eqnarray}
Hereafter, we will ignore the $-\left({e}/{m_{0}c}\right)\bm{A}(t) \cdot \hbar \bm{k}$ term because it is a classical number and only causes a shift in the total energy. 
Thus, we can derive the total effective Hamiltonian $H_{\bm{k}}^{\rm{eff}}$ as follows:
\begin{eqnarray}
H^{\rm{eff}}_{\bm{k}} &=& H^{\rm{eff}}_{0,\bm{k}} + H^{\rm{eff}}_{I,\bm{k}}, 
\label{eq:Heff1}
\\ 
H^{\rm{eff}}_{0,\bm{k}} &=& \frac{{\bm {p}}^2}{2 m_{0}} + \sum_{i}V(\bm{x}-\bm{R}_{i}) + \frac{\hbar}{m_{0}}{\bm{k}} \cdot {\bm{p}} + \frac{\hbar^2 \bm{k}^2}{2 m_{0}}, 
\label{eq:Heff2}
\\
H^{\rm{eff}}_{I,\bm{k}} &=& -\frac{e}{m_{0}c} \bm{A}(t) \cdot {\bm{p}}.
\label{eq:Heff3}
\end{eqnarray}

Below, we focus on the Luttinger-Kohn model\cite{Luttinger_S, Chuang_S, Ahn_S, Dargys_S, Pryor_S, Luque_S, Tomic_S, Sytnyk_S, Bastos_S, Pfeffer_S} where the effective Hamiltonian $H^{\rm eff}_{0,\bm{k}}$ can be expressed as an 8$\times$8 matrix. 
Each element is related to the conduction, heavy-hole, light-hole, and split-off bands for spin up and spin down components. 
The basis set of the Luttinger-Kohn model is the angular momentum basis $\ket{J,{J}_{z}}$, where
\begin{eqnarray}
&\ket{u_{1}}& \equiv \ket{\frac{1}{2},+\frac{1}{2}} = \ket{s; \uparrow}, \\
&\ket{u_{2}}& \equiv \ket{\frac{3}{2},+\frac{3}{2}} = \frac{i}{\sqrt{2}} \left[\ket{x; \uparrow} + i\ket{y; \uparrow}\right],\\
&\ket{u_{3}}& \equiv \ket{\frac{3}{2},+\frac{1}{2}} = \frac{i}{\sqrt{6}} \left[\ket{x; \downarrow} + i\ket{y; \downarrow} -2 \ket{z; \uparrow}\right], \\
&\ket{u_{4}}& \equiv \ket{\frac{1}{2},+\frac{1}{2}} = \frac{i}{\sqrt{3}} \left[\ket{x; \downarrow} + i\ket{y; \downarrow} + \ket{z; \uparrow}\right]. 
\end{eqnarray}
Here, we define $\ket{s;\sigma}$, $\ket{x;\sigma}$, $\ket{y;\sigma}$, and $\ket{z;\sigma}$ to be the $s$, $p_{x}$, $p_{y}$, and $p_{z}$ wavefunctions, respectively, for spin components $\sigma = \uparrow$ or $\downarrow$. 
The remaining set of Bloch basis states are expressed as
\begin{eqnarray}
&\ket{u_{5}}& \equiv \ket{\frac{1}{2},-\frac{1}{2}} = -\ket{s; \downarrow}, \\
&\ket{u_{6}}& \equiv \ket{\frac{3}{2},-\frac{3}{2}} = -\frac{i}{\sqrt{2}} \left[\ket{x; \downarrow} - i\ket{y; \downarrow}\right],\\
&\ket{u_{7}}& \equiv \ket{\frac{3}{2},-\frac{1}{2}} = \frac{i}{\sqrt{6}} \left[\ket{x; \uparrow} - i\ket{y; \uparrow} +2 \ket{z; \downarrow}\right], \\
&\ket{u_{8}}& \equiv \ket{\frac{1}{2},-\frac{1}{2}} = \frac{i}{\sqrt{3}} \left[\ket{x; \uparrow} - i\ket{y; \uparrow} - \ket{z; \downarrow}\right]. 
\label{basis2}
\end{eqnarray}
Supposing the above basis, we can describe the 8-band $\bm{k} \cdot \bm{p}$ Hamiltonian as follows:
\begin{eqnarray}
H^{\rm{eff}}_{0,\bm{k}} = \left( \begin{array}{cc} 
H^{\bm{k}}_{uu} & H^{\bm{k}}_{ul} \\ H^{\bm{k}}_{lu} & H^{\bm{k}}_{ll} \\
\end{array} \right).
\label{Hamiltonian0}
\end{eqnarray}
Here, $H^{\bm{k}}_{uu}$, $H^{\bm{k}}_{ul}$, $H^{\bm{k}}_{lu}$, and $H^{\bm{k}}_{ll}$ are 4 $\times$ 4 submatrices. 
{The submatrix $H^{\bm{k}}_{uu}$} has the form\cite{Tomic_S},
\begin{eqnarray}
H^{\bm{k}}_{uu} = \left( \begin{array}{cccc} 
E^{\bm{k}}_{CB} & -\sqrt{3}T_{\bm{k}} & \sqrt{2}U_{\bm{k}} & -U_{\bm{k}} \\ -\sqrt{3}T_{\bm{k}}^{*} & E^{\bm{k}}_{HH} & \sqrt{2}S_{\bm{k}} & -S_{\bm{k}} \\ \sqrt{2}U_{\bm{k}} & \sqrt{2}S_{\bm{k}}^{*} & E^{\bm{k}}_{LH} & -\sqrt{2}Q_{\bm{k}} \\ -U_{\bm{k}} & -S_{\bm{k}}^{*} & -\sqrt{2}Q_{\bm{k}} & E^{\bm{k}}_{SO}\\
\end{array} \right),
\label{Hamiltonian1}
\end{eqnarray}
while the submatrix $H^{\bm{k}}_{ll}$ is $H^{\bm{k}}_{ll}=H_{uu}^{\bm{k}*}$. 
The submatrices, $H^{\bm{k}}_{ul}$ and $H^{\bm{k}}_{lu}$, are expressed as
\begin{eqnarray}
H^{\bm{k}}_{ul} = \left( \begin{array}{cccc} 
0 & 0 & -T_{\bm{k}}^{*} & -\sqrt{2}T_{\bm{k}}^{*} \\ 0 & 0 & -R_{\bm{k}} & -\sqrt{2}R_{\bm{k}} \\ T_{\bm{k}}^{*} & R_{\bm{k}} & 0 & \sqrt{3}S_{\bm{k}} \\ \sqrt{2}T_{\bm{k}}^{*} & \sqrt{2}R_{\bm{k}} & -\sqrt{3}S_{\bm{k}} & 0\\
\end{array} \right)
\end{eqnarray}
and $H^{\bm{k}}_{lu}=H_{ul}^{\bm{k}*}$. 
The diagonal elements of $H^{\bm{k}}_{uu}$ and $H^{\bm{k}}_{ll}$ are defined as
\begin{eqnarray}
&E^{\bm{k}}_{CB}& = E_{g} + O_{\bm{k}}, \\
&E^{\bm{k}}_{HH}& = - \left(P_{\bm{k}} + Q_{\bm{k}}\right), \\
&E^{\bm{k}}_{LH}& = - \left(P_{\bm{k}} - Q_{\bm{k}}\right), \\
&E^{\bm{k}}_{SO}& = - \left(P_{\bm{k}} + \Delta_{SO}\right). 
\label{eq:Hamiltonian3}
\end{eqnarray}
The subscripts CB, HH, LH, and SO stand for conduction, heavy-hole, light-hole and split-off bands, respectively, and $\Delta_{SO}$ and $E_{g}$ are the split-off energy and the band-gap energy. 
Other Hamiltonian matrix elements are given {as}
\begin{eqnarray}
&O_{\bm{k}}& = \frac{\hbar^2}{2 m_{0}}\gamma_{C}\left(k_{x}^2 + {k}_{y}^2 + {k}_{z}^2 \right), \\
&P_{\bm{k}}& = \frac{\hbar^2}{2 m_{0}}\gamma_{1}\left(k_{x}^2 + {k}_{y}^2 + {k}_{z}^2 \right), \\
&Q_{\bm{k}}& = \frac{\hbar^2}{2 m_{0}}\gamma_{2}\left(k_{x}^2 + k_{y}^2 - 2 k_{z}^2 \right), \\
&R_{\bm{k}}& = \frac{\hbar^2}{2 m_{0}}\sqrt{3}\left[\gamma_{2}(k_{x}^2 - k_{y}^2) - 2 i \gamma_{3} k_{x} k_{y} \right], \\
&S_{\bm{k}}& = \frac{\hbar^2}{2 m_{0}}\sqrt{6}\gamma_{3}\left(k_{x} - i k_{y}\right)k_{z}, \\
&T_{\bm{k}}& = \frac{1}{\sqrt{6}}P_{0}\left(k_{x} + i k_{y}\right), \\
&U_{\bm{k}}& = \frac{1}{\sqrt{3}}P_{0} k_{z}. 
\label{eq:Hamiltonian4}
\end{eqnarray}
Here, $k_{x}$, $k_{y}$, and $k_{z}$ denote components of the Bloch wavevector along the [100], [010], and [001] crystallographic directions, respectively, and $\gamma_{0}$, $\gamma_{1}$, $\gamma_{2}$, and $\gamma_{3}$ are the Luttinger parameters.
We set the Luttinger parameters of GaAs to be $\gamma_{C}=0.5$, $\gamma_{1}=2.7$, $\gamma_{2}=-0.1$, and $\gamma_{3}=0.7$ {following Ref. \onlinecite{Sytnyk_S}}. 
The dipole matrix element (the Kane matrix element) is defined as
\begin{equation}
P_{0} = -i(\hbar/m_{0})\braket{s;\sigma|p_{z}|z;\sigma}. 
\end{equation}
Diagonalization of this Hamiltonian yields the eight-band structure of the GaAs near the {$\Gamma$} point.

To express the light-matter interaction in a simple form, we will return to the elemental basis set of $\ket{s;\uparrow}$, $\ket{x;\uparrow}$, $\ket{y;\uparrow}$, $\ket{z;\uparrow}$, $\ket{s;\downarrow}$, $\ket{x;\downarrow}$, $\ket{y;\downarrow}$, and $\ket{z;\downarrow}$. 
A linear combination of this basis set constructs the conventional basis of GaAs near the $\Gamma$ point, $\ket{u_{i}}$, and a unitary transformation to the elementary basis can be expressed by a matrix whose elements are $\braket{u_{i}|v;\sigma}$. 
Here, $\sigma$ is the spin index of the wavefunctions, and $v$ is the index of $s$, $p_{x}, p_{y}$, and $p_{z}$. 
By performing simple analytic calculations, we can express the unitary matrix $U$ in the following form:
\begin{eqnarray}
U&=&\left( \begin{array}{cccccccc}
1 & 0 & 0 & 0 & 0 & 0 & 0 & 0 \\ 0 & -\frac{i}{\sqrt{2}} & 0 & 0 & 0 & 0 & -\frac{i}{\sqrt{6}} & -\frac{i}{\sqrt{3}} \\ 0 & -\frac{1}{\sqrt{2}} & 0 & 0 & 0 & 0 & \frac{1}{\sqrt{6}} & \frac{1}{\sqrt{3}} \\ 0 & 0 & i\sqrt{\frac{2}{3}} & -\frac{i}{\sqrt{3}} & 0 & 0 & 0 & 0 \\
0 & 0 & 0 & 0 & -1 & 0 & 0 & 0 \\ 0 & 0 & -\frac{i}{\sqrt{6}} & -\frac{i}{\sqrt{3}} & 0 & \frac{i}{\sqrt{2}} & 0 & 0 \\ 0 & 0 & -\frac{1}{\sqrt{6}} & -\frac{1}{\sqrt{3}} & 0 & -\frac{1}{\sqrt{2}} & 0 & 0 \\ 0 & 0 & 0 & 0 & 0 & 0 & -\sqrt{\frac{2}{3}}i & \frac{i}{\sqrt{3}} \\
\end{array} \right). 
\end{eqnarray}
By performing the unitary transformation $H^{\rm{eff}'}_{\bm{k}}=U^{\dagger}H^{\rm{eff}}_{0,\bm{k}}U $, we can obtain the Hamiltonian in the new basis. 
{This $8\times 8$ matrix $H^{\rm{eff}'}_{\bm{k}}$ corresponds to $H_{\bm k}$, given in Eq. (1) in the main text.}

Next, let us consider the light-matter interaction. 
As derived above, the effective light-matter interaction Hamiltonian is expressed by $H^{\rm{eff}}_{I,{\bm{k}}}= -({e}/{m_{0}c}) \bm{A}(t) \cdot {\bm{p}}$. 
Supposing the vector potential ${\bm{A}}(t)$ to be ${\bm{A}}(t) = (0,0,A_{z}(t))$, we can derive the light-matter interaction Hamiltonian as $H_{I,{\bm{k}}}^{\rm{eff}} = -(e/m_{0}c) A_{z}(t) p_{z}$. 
In the basis set of $\ket{s;\sigma}$, $\ket{x;\sigma}$, $\ket{y;\sigma}$, and $\ket{z;\sigma}$, almost all of the matrix elements in $H^{\rm{eff}}_{I,\bm{k}}$ becomes zero because of the parity symmetry. 
The nonzero matrix elements take the form,
\begin{eqnarray}
\braket{s;\sigma|H^{\rm{eff}}_{I,\bm{k}}|z;\sigma} &=& -\frac{e}{m_{0}{c}}A_{z}\braket{s;\sigma|p_{z}|z;\sigma} = -i \hbar \Omega_{R}(t){,}
\end{eqnarray}
or its complex conjugate, where we have defined the Rabi frequency as $\Omega_R(t) = (e/c\hbar^{2})A_{z}P_{0}$. 
Thus, we obtain {the} following matrix form of $H_{I,{\bm{k}}}^{\rm{eff'}}$:
\begin{eqnarray}
H_{I,{\bm{k}}}^{\rm{eff}'}&=&\hbar \left( \begin{array}{cccccccc}
0 & 0 & 0 & -i \Omega_{R}(t) & 0 & 0 & 0 & 0 \\ 0 & 0 & 0 & 0 & 0 & 0 & 0 & 0 \\ 0 & 0 & 0 & 0 & 0 & 0 & 0 & 0 \\ i \Omega_{R}^{*}(t) & 0 & 0 & 0 & 0 & 0 & 0 & 0 \\
0 & 0 & 0 & 0 & 0 & 0 & 0 & -i \Omega_{R}(t) \\ 0 & 0 & 0 & 0 & 0 & 0 & 0 & 0 \\ 0 & 0 & 0 & 0 & 0 & 0 & 0 & 0 \\ 0 & 0 & 0 & 0 & i \Omega_{R}^{*}(t) & 0 & 0 & 0 \\
\end{array} \right) 
\end{eqnarray}
The total Hamiltonian is thus $H_{{\bm{k}}}^{\rm{eff}'}=H^{\rm{eff}'}_{0,{\bm{k}}}+H^{\rm{eff}'}_{I,{\bm{k}}}$. 
The time evolutions of the wavefunction, $\ket{\psi}_{{\bm{k}}}=(\psi^{{\bm{k}}}_{s\uparrow},\psi^{{\bm{k}}}_{x\uparrow},\psi^{{\bm{k}}}_{y\uparrow},\psi^{{\bm{k}}}_{z\uparrow},\psi^{{\bm{k}}}_{s\downarrow},\psi^{{\bm{k}}}_{x\downarrow},\psi^{{\bm{k}}}_{y\downarrow},\psi^{{\bm{k}}}_{z\downarrow})^{\dagger}$, can be obtained by solving the Schr\"odinger eq{u}ation {for each wavenumber ${\bm k}$}, where 
\begin{eqnarray}
i \hbar \frac{\partial}{\partial t}\ket{\psi_{{\bm{k}}}} = H_{{\bm{k}}}^{\rm{eff}'} \ket{\psi_{\bm{k}}}
\end{eqnarray}
On the basis of these solutions, the generated current can be calculated from the following definitions:
\begin{eqnarray}
J_{z}(t) = -\frac{1}{c}\left< \frac{\partial H_{I}}{\partial A_{z}}\right> \propto \sum_{\bm{k}}-i\left[\psi_{s\uparrow}^{{\bm{k}}*}(t)\psi^{\bm{k}}_{z\uparrow}(t) - \psi_{z\uparrow}^{{\bm{k}}*}(t)\psi^{\bm{k}}_{s\uparrow}(t) + \psi_{s\downarrow}^{{\bm{k}}*}(t)\psi^{\bm{k}}_{z\downarrow}(t) - \psi_{z\downarrow}^{{\bm{k}}*}(t)\psi^{\bm{k}}_{s\downarrow}(t)\right]. \nonumber
\end{eqnarray}
Finally, the HHG spectra in GaAs {is calculated} as $I=\left|\omega {\cal J}_{z}(\omega) \right|^2$, where ${\cal J}_{z}(\omega)$ is the Fourier transform of the generated current {$J_{z}(t)$}. 

\begin{figure}[t]
\includegraphics[width= 100mm]{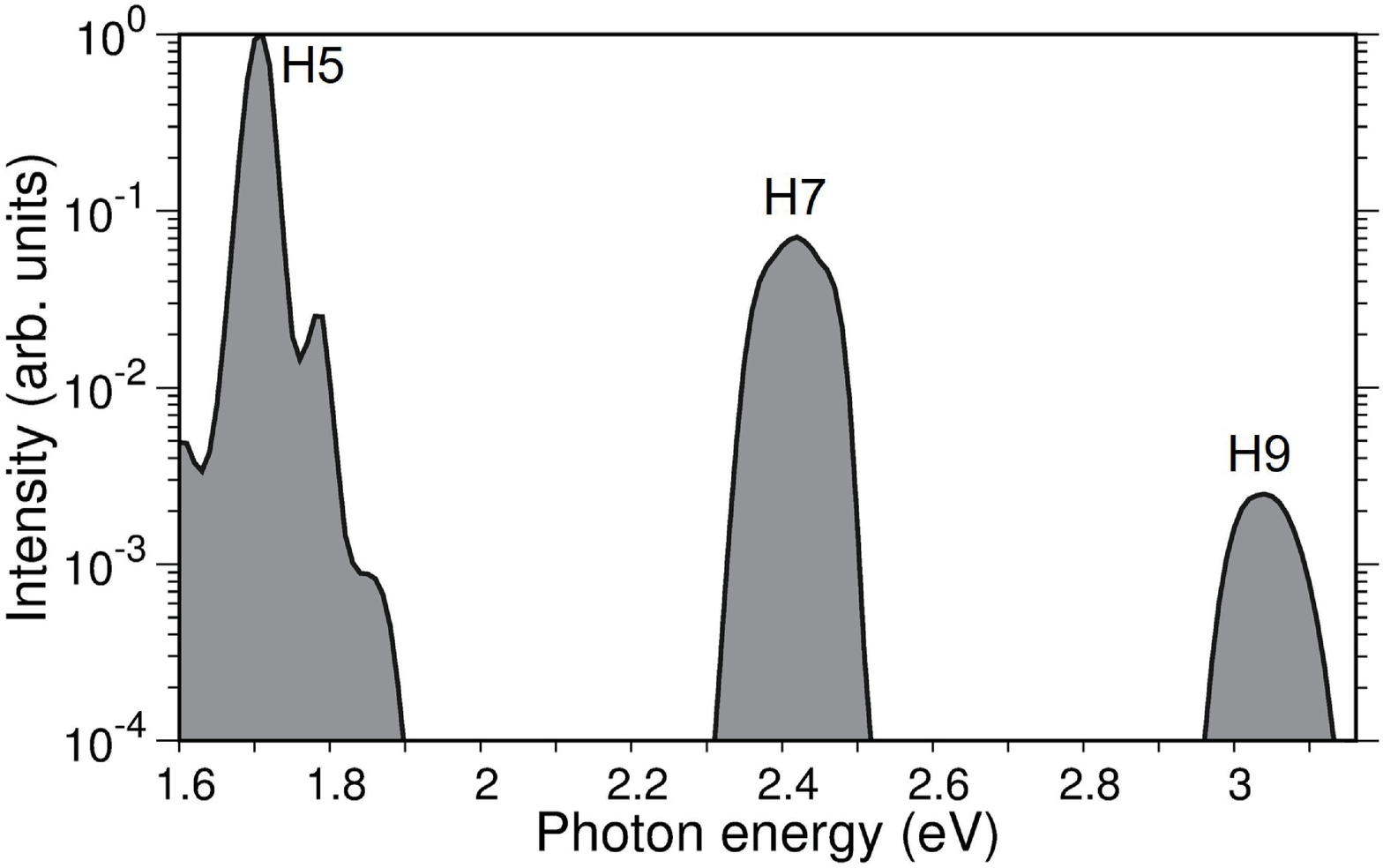}
\caption{Example of HHG spectrum calculated from the Luttinger-Kohn model.}
\label{figS0}
\end{figure}

The initial conditions of the system are supposed to be the occupied valence bands. 
{W}e performed numerical calculations starting from an initial state in which one of the six valence bands is occupied. 
We calculated the time evolutions of the generated current by summing all the solutions obtained for the six initial conditions.

Figure \ref{figS0} shows an example of an HHG spectrum calculated from the above model. 
The center photon energies of the harmonics correspond closely to the odd multiples of the excitation photon energy (0.34 eV), while each harmonic has a specific spectral shape that is consistent with the experimental results (see Fig. 1(d) in the main text).
We could identify that these shapes are easily changed by changing the excitation waveform, which is also consistent with the experimental results (see Section V).

\section{Theoretical Analysis based on the Kane model}
We showed in the main text that the intensities of 5th, 7th, and 9th HHG in the non-perturbative regime have dips as a function of the field strength. 
To understand the origin of these dips, we constructed a Kane model with a conduction band and a split-off band, where the model parameters are related to the ones in the Kohn-Luttinger model described in the previous section. 
Using the Kane model, we will discuss three features of the nonmonotonic behavior of the HHG intensities. 
First, we show that the nonmonotonic behavior appears even in the Kane model. 
This indicates that the number of valence bands is not essential to the appearance of the dips. 
Second, we clarify that the time-dependent change in the band energies induced by the intra-band transition (field-induced dynamic band structures) affects the dips. 
Third, we roughly estimate the intensities of the external electric field at which the HHG intensities deviate from the predictions of perturbation theory and compare these intensities with the experimental results.

Let us regard ${(\hbar}/{m_{0}}){\bm{k}} \cdot {\bm{p}}$ in the Hamiltonian (\ref{eq:Heff2}) as a pertur{b}ation. 
{Using} the eight {bases} defined in Eqs. (9)-(16) and {employing} { conventional} pertur{b}ation theory, we can derive the eigenvalues of the conduction $E^{\sigma}_{c}$ and the {split}-off bands $E^{\sigma}_{v}$ as follows:
\begin{eqnarray}
&E^{\sigma}_{c}& = \frac{\hbar^2 \bm{k}^2}{2 m_{0}}+\frac{P_{0}^{2}}{E_g}\left[k^2_{x} + k^2_{y} + k^2_{z}\right] + E_g , \\
&E^{\sigma}_{v}& = -\frac{\hbar^2 \bm{k}^2}{2 m_{0}}-\frac{P_{0}^{2}}{3 E_{g}}\left[k^2_{x} + k^2_{y} + k^2_{z}\right], 
\end{eqnarray}
where we {have assumed} $E_{g} \gg \Delta_{SO}$. 
We express the Bloch wavefunctions of the conduction band $\ket{1/2,\pm 1/2}'_c$ and the split-off band $\ket{1/2,\pm 1/2}'_v$ as
\begin{eqnarray}
&\ket{\frac{1}{2},+\frac{1}{2}}'_c& = \ket{s; \uparrow} - i \alpha \left[k_{x} \ket{x; \uparrow} + k_{y} \ket{y; \uparrow} + k_{z} \ket{z; \uparrow} \right], \\
&\ket{\frac{1}{2},-\frac{1}{2}}'_c& = -\ket{s; \downarrow} + i \alpha \left[k_{x} \ket{x; \downarrow} + k_{y} \ket{y; \downarrow} + k_{z} \ket{z; \downarrow} \right], \\
&\ket{\frac{1}{2},+\frac{1}{2}}'_v& = \frac{i}{\sqrt{3}}\left(\ket{x; \downarrow} + i\ket{y; \downarrow} +\ket{z; \uparrow} \right) \nonumber \\ 
&& \hspace{15mm} + \frac{1}{\sqrt{3}} \alpha k_{z} \ket{s; \uparrow} + \frac{1}{\sqrt{3}} \alpha (k_{x} - i k_{y}) \ket{s; \downarrow} \\
&\ket{\frac{1}{2},-\frac{1}{2}}'_v& = \frac{i}{\sqrt{3}}\left(\ket{x; \uparrow} - i\ket{y; \uparrow} -\ket{z; \downarrow} \right) \nonumber \\ 
&& \hspace{15mm} + \frac{1}{\sqrt{3}} \alpha (k_{x} + i k_{y}) \ket{s; \uparrow} - \frac{1}{\sqrt{3}} \alpha k_{z} \ket{s; \downarrow},
\end{eqnarray}
where we have define{d} $\alpha= P_{0}/E^{\rm{Re}}_{g}$ and $E^{\rm{Re}}_{g}=E_g + \Delta_{SO}$. 
Hereafter, we only focus on the{se} four bands, i.e., the conduction {bands} and split-off bands for the spin up and down components. 

Now, let us assume the vector potential ${\bm{A}}(t)$ to be ${\bm{A}}(t) = (0,0,A_{z}(t))$, i.e., $H_{I,{\bm{k}}}^{\rm{eff}} = -(e/m_{0}c) A_{z}(t) p_{z}$. 
With the four bases defined above, we can derive the single-particle Hamiltonian $H^{\rm{eff}}_{0,\bm{k}}$ as
\begin{eqnarray}
H^{\rm{eff}}_{0,\bm{k}} = \left( \begin{array}{cccc} 
E^{\uparrow}_{c}(\bm{k}) & 0 & 0 & 0 \\ 0 & E^{\downarrow}_{c}(\bm{k}) & 0 & 0 \\ 0 & 0 & E^{\uparrow}_{v}(\bm{k}) & 0 \\ 0 & 0 & 0 & E^{\downarrow}_{v}(\bm{k})\\
\end{array} \right){,}
\label{Hamiltonian10}
\end{eqnarray}
where $E^{\sigma}_{c}$ and $E^{\sigma}_{v}$ are defined in Eqs. (36) and (37). 
Similarly, we can derive a light-matter interaction Hamiltonian $H^{\rm{eff}}_{I,\bm{k}}$ of the form:
\begin{eqnarray}
H^{\rm{eff}}_{I,\bm{k}} = \hbar \Omega_{R}(t) \left(\begin{array}{cccc} 
-{2} \alpha k_{z} & 0 & L & J \\ 0 & -{2 \alpha} k_{z} & -J^{*} & L \\ L^{*} & -J & 
\frac{2}{3} \alpha k_{z} & 0 \\ J^{*} & L^{*} & 0 & \frac{2}{3} \alpha k_{z} \\
\end{array} \right),
\label{Hamiltonian20}
\end{eqnarray}
where $L = -({1}/{\sqrt{3}})[1 - \alpha^2 k_{z}^{2}]$ and $J = ({1}/{\sqrt{3}}) \alpha^{2} k_{z} (k_{x} + i k_{y})$. 
{The Hamiltonian of the Kane model in Eq. (3) of the main text is obtained if we drop the terms proportional to $\alpha^2$ under the assumption that $\alpha$ is small. 
We find that this approximate Hamiltonian gives almost the same results for the parameters used in the present calculation. 
In the following calculation, we do not drop these terms} and use the expression ${\Omega_{R}(t)=\Omega_{R0} \exp\left[-(t-t_{0})^2/T^2 \right] \cos \omega t}$, where $\Omega_{R0} = d \cdot E_{0}/ \hbar$.
Thus, the matrix of the total Hamiltonian $H_{{\bm{k}}}^{\rm{eff}}=H^{\rm{eff}}_{0,{\bm{k}}}+H^{\rm{eff}}_{I,{\bm{k}}}$ can be derived. 
The time evolutions of the wavefunction, $\ket{\psi_{{\bm{k}}}} = \left(\psi_{c\uparrow}({\bm{k}}),\psi_{c\downarrow}({\bm{k}}),\psi_{v\uparrow}({\bm{k}}),\psi_{v\downarrow}({\bm{k}})\right)^{\dagger}$, can be obtained by solving the Schr\"odinger eqation $i \hbar \frac{\partial}{\partial t}\ket{\psi_{{\bm{k}}}} = H_{{\bm{k}}}^{\rm{eff}} \ket{\psi_{\bm{k}}}$. 
The explicit form of these equations for up and down spin{s} is as follows:
\begin{eqnarray}
i \hbar \dot{\psi}_{c\sigma}({\bm{k}}) &=& \left[E_{c\sigma}({\bm{k}}) -2 \hbar \Omega_{R}(t) \alpha k_{z} \right]\psi_{c\sigma}({\bm{k}}) \nonumber \\ 
&&- \frac{1}{\sqrt{3}} \hbar \Omega_{R}(t) \left[1 - \alpha^2 k_{z}^{2}\right]\psi_{v\sigma}({\bm{k}}) \pm \frac{1}{\sqrt{3}} \hbar \Omega_{R}(t) \alpha^{2} k_{z} (k_{x} \pm i k_{y})\psi_{v\sigma'}({{\bm{k}}}), \\
i \hbar \dot{\psi}_{v\sigma}({\bm{k}}) &=& -\frac{1}{\sqrt{3}} \hbar \Omega_{R}(t) \left[1 - \alpha^2 k_{z}^{2}\right]\psi_{c\sigma}({\bm{k}}) \nonumber \\ 
&& \mp \frac{1}{\sqrt{3}} \hbar \Omega_{R}(t)\alpha^{2} k_{z} (k_{x} \pm i k_{y})\psi_{c\sigma'}({{\bm{k}}}) +\left[E_{v\sigma}({\bm{k}}) + \frac{2}{3} \hbar \Omega_{R}(t) \alpha k_{z} \right]\psi_{v\sigma}({\bm{k}}).
\label{eq:timeE}
\end{eqnarray}
Here, $\sigma$ and $\sigma'$ are opposite spin indices, and the $\pm$ sign corresponds to $\sigma=\uparrow$ and $\downarrow$, respectively. 
By solving these equations, we obtain the time evolution of the wavefunctions $\psi_{c\uparrow}({\bm{k}})$, $\psi_{c\downarrow}({\bm{k}})$, $\psi_{v\uparrow}({\bm{k}})$, and $\psi_{v\downarrow}(\bm{k})$. 
{After that,} we calculate the generated current, {defined as}
\begin{eqnarray}
J_{z}(t) &=& -\frac{1}{c}\left< \frac{\partial H_{I}}{\partial A_{z}}\right> \propto \sum_{\bm{k},\sigma} \alpha k_{z} \left[-2 |\psi_{c\sigma}({\bm{k}})|^{2} +\frac{2}{3} |\psi_{v\sigma}({\bm{k}})|^{2} \right] \nonumber \\
&+& \sum_{\bm{k},\sigma}\frac{2}{\sqrt{3}} {\rm{Re}}\left[(1 - \alpha^{2} k_{z}^{2})\psi_{c\sigma}({{\bm{k}}})\psi_{v\sigma}^{*}({{\bm{k}}})\right]- \sum_{\bm{k},\sigma}\frac{2}{\sqrt{3}}{\rm{Re}}\left[\alpha^{2} k_{z} (k_{x} \pm i k_{y})\psi_{c\sigma}({{\bm{k}}})\psi_{v\sigma'}^{*}({{\bm{k}}})\right]. \nonumber
\label{eq:Current}
\end{eqnarray}
Here, $\pm$ sign corresponds to $(\sigma,\sigma') = (\uparrow,\downarrow)$ or $(\downarrow,\uparrow)$. 
Finally, the HHG spectra in GaAs are calculated as $I=\left|\omega {\cal J}_{z}(\omega) \right|^2$, where ${\cal J}_{z}(\omega)$ is the Fourier transform of the generated current along the $z$-axis. 
In this numerical calculation, we set the band-gap energy and the dipole moment of GaAs to $E_g = 4.2 \hbar \omega$ and $d = 1.8 \ [\rm{e} \cdot \rm{nm}]$, respectively.
The dipole moment is slightly larger compared to the conventional ones. 
We consider this difference to be derived from an approximate estimation of the light source intensity.

To examine the origin of the dips in the HHG intensities, we focus on the light-matter interaction described by Eq. (\ref{Hamiltonian20}) and separate it into two contributions as follows:
\begin{eqnarray}
H^{\rm{eff}}_{I,\bm{k}} &=& H^{\rm{eff}}_{{\rm intra},\bm{k}} + H^{\rm{eff}}_{{\rm inter},\bm{k}}, \\
H^{\rm{eff}}_{{\rm intra},\bm{k}} &=&
\hbar \Omega_{R}(t) \left(\begin{array}{cccc} 
-2 \alpha k_{z} & 0 & 0 & 0 \\ 0 & -2 \alpha k_{z} & 0 & 0 \\ 0 & 0 & \frac{2}{3} \alpha k_{z} & 0 \\ 0 & 0 & 0 & \frac{2}{3} \alpha k_{z} 
\end{array} \right), 
\label{eq:Hintra} \\
H^{\rm{eff}}_{{\rm inter},\bm{k}} &=&
\hbar \Omega_{R}(t) \left(\begin{array}{cccc} 
0 & 0 & L & J \\ 0 & 0 & -J^{*} & L \\ L^{*} & -J & 0 & 0 \\ J^{*} & L^{*} & 0 & 0 
\end{array} \right).
\end{eqnarray}
Here, $H^{\rm{eff}}_{{\rm intra},{\bm k}}$ temporally changes the band energies, while $H^{\rm{eff}}_{{\rm inter},{\bm k}}$ describes transitions between the bands. 
We call the former (latter) part the intraband (interband) contribution of the light-matter interaction Hamiltonian. 
Figures 3(c) and (d) in the main text indicate the HHG intensities calculated for the Kane model with and without the intraband contribution, respectively. 
These figures show that the dips become less significant if the intraband contribution is neglected. 
This result points out the importance of the field-induced dynamic band structures, i.e., the intraband contribution of the matter-light interaction.

Figure 2(a) and (b) in the main text also indicates that the HHG intensities initially follow the scaling laws of nonlinear optics in the weak-intensity perturbative regime, before deviating from them as the incident electric-field intensity increases. 
Let us first estimate the intensity of the electric field at which the HHG intensities deviate from perturbation theory since this roughly gives the electric field at which the nonmonotonic behavior starts to appear. 
In our previous work \cite{Tamaya2019PRBR_S}, we showed that the intraband contribution, i.e., the {field-induced dynamic band structures} can be understood within the Floquet subband picture. 
The weight of the $n$-th subband for conduction electrons is estimated as $J_n(A/\hbar \omega)$, where $A \simeq 2\hbar \Omega_{R0} \alpha k_z$ is the maximum energy shift of the conduction band due to the matter-light interaction (see Eq. (\ref{eq:Hintra})). 
The nonperturbative effect becomes significant when the weight of the first subband, $J_1(A/\hbar \omega)$, becomes of order of 1. 
This condition is roughly $2 \Omega_{R0} \alpha k_{z}/ \omega \approx 1$, i.e., $2 \tilde{\Omega}_{R0} \tilde{k}_{z}/\tilde{E}^{\rm{Re}}_{g} \approx 1$, where $\tilde{k}_{z} = k_{z} P_{0}/\hbar \omega$ and $\tilde{E}^{\rm{Re}}_{g} = E^{\rm{Re}}_{g}/\hbar \omega$. 
If we roughly estimate the typical wavenumber as $\tilde{k}_{z} \approx 1.2$ from the cutoff wavenumber and $\tilde{E}^{\rm{Re}}_{g} = 5.2$, we can derive the crossover intensity as $E \approx 4 \rm{MV/cm}$, which is consistent with the numerical results for both the Kane model (Fig. 3(c)) and the Luttinger model (Fig. 2(b)), as well as the experimental results in the main text (Fig. 2(a)).

\section{Spatial nonuniformity in the field-intensity dependence}
\begin{figure}[t]
\includegraphics[width=130mm]{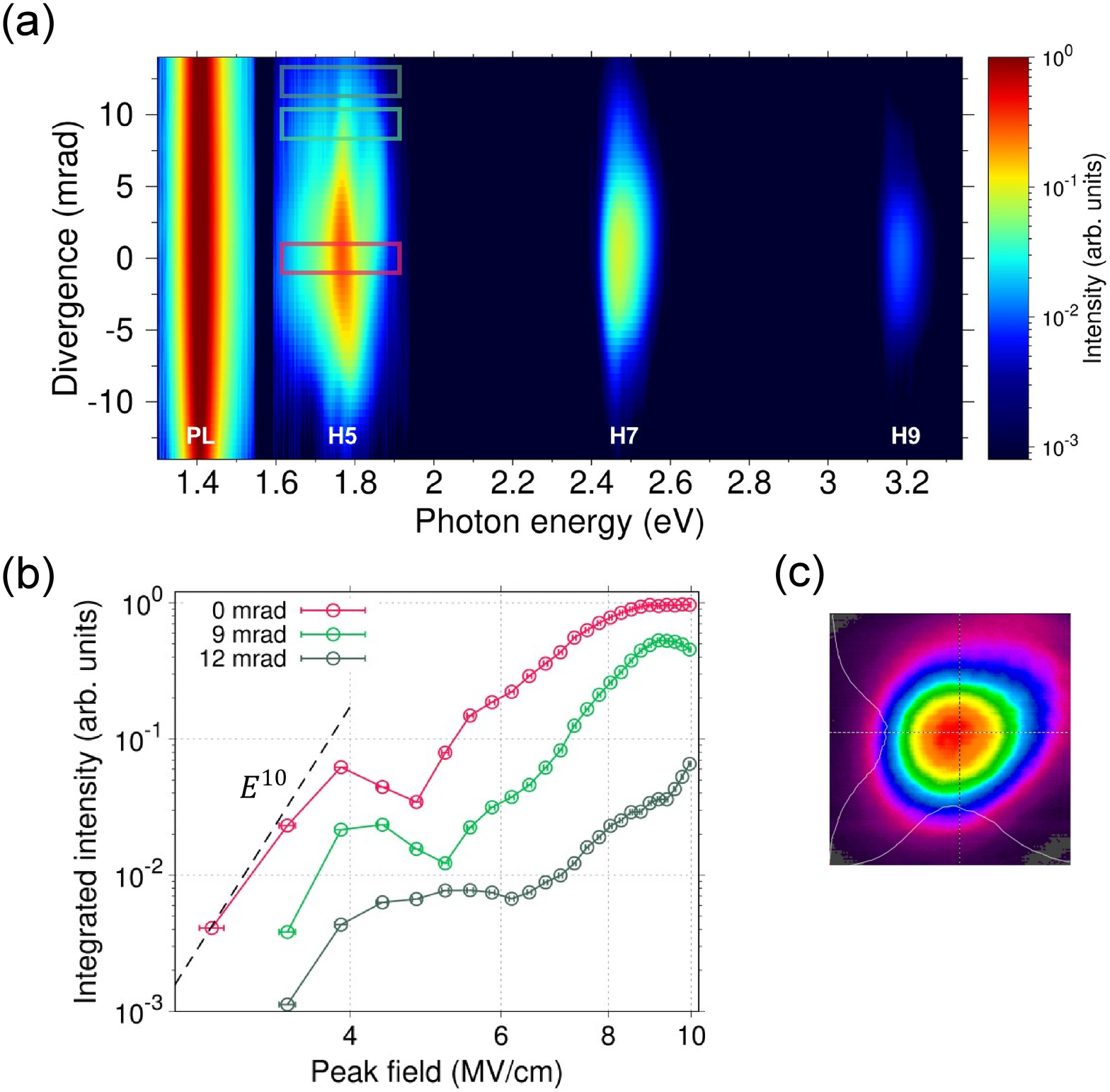}
\caption{ (a) Horizontal profiles of HHG and photoluminescence (PL) spectra in GaAs at peak field of 10 MV/cm. (b) Field-intensity dependences of the 5th harmonic intensity at the different transverse positions indicated in (a). (c) Beam profile of mid-IR laser source before the focus lens. }
\label{figS5}
\end{figure}

The macroscopic properties of HHG have been experimentally and theoretically shown to have spatial nonuniformity caused by the transverse intensity distribution of the drive laser \cite{Floss2018_S,Abadie2018_S,Zair2008_S,Toma1999_S}. To study the field-intensity dependence of HHG in gaseous media, Za\"{\i}r {\sl et al.}\cite{Zair2008_S} demonstrated that the spatial averaging effect could be partly avoided by employing far-field spatial filtering, and Toma {\sl et al.}\cite{Toma1999_S} demonstrated spatial shaping of the focused beam with a flat-top intensity profile.

Here, we characterized the spatial nonuniformity in the field-intensity dependence of HHG in GaAs in reflection geometry. We used a linearly polarized 60-fs mid-IR laser source at 3.5 $\mu$m with a near-Gaussian transverse profile, as shown in Fig. \ref{figS5}(c). The linear polarization was along the [001] axis of GaAs. The mid-IR pulses were focused with a lens having an $f$-number of $f$/32 ($\sim$31 mrad) on the (110) surface of GaAs, and the reflected beam was collimated by a concave mirror ($R=100$ mm). We inserted an iris in the collimated beam path and measured HHG spectra as a function of the horizontal position of the iris, while the vertical position was fixed to the vertical center point of the beam. Figure \ref{figS5}(a) shows that the beam profile of HHG had the highest intensity in the center, while the background fluorescence around 1.42 eV of GaAs was almost uniform. In comparison with the 5th harmonic intensity as a function of the peak field intensity in Fig \ref{figS5}(b), we found that the oscillatory behavior varied with the iris position. The first dips in the field-intensity dependences appeared at $\sim$4.5, $\sim$5, and $\sim$6.5 MV/cm for the 0, 9, and 12 mrad positions, respectively. From these results, we considered that the effective {driving} field intensity was lower in the outer part of the HHG beam. Therefore, we detected the center of the beam to avoid spatial averaging of the dips {in addition to minimizing} the signal level of background fluorescence.

\section{Comparison of HHG in transmission geometry}
\begin{figure}[t]
\includegraphics[width=150mm]{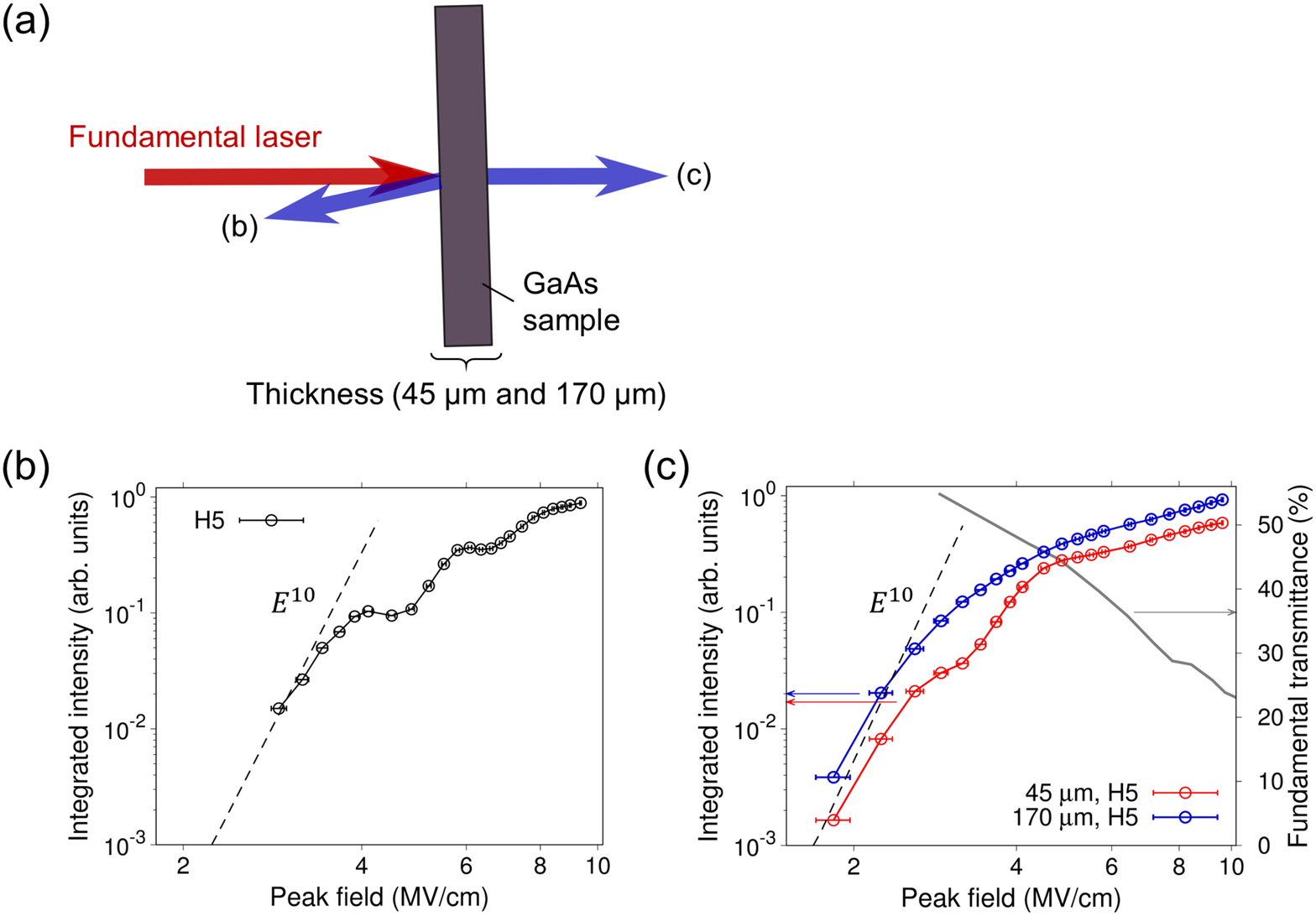}
\caption{(a) Schematic diagram of HHG in reflection and transmission geometry. (b) Field-intensity dependence of the 5th harmonic intensity in reflection geometry and (c) in transmission geometry. The gray line shows the transmittance of the mid-IR pulses in the 170-$\mu$m-thick sample measured as a function of incident laser intensity. The transmittance in the weak-field limit is estimated to be 56\% from the Fresnel formula. }
\label{figS6}
\end{figure}

Figure \ref{figS6} demonstrates that the field-intensity dependence in transmission geometry significantly changed from that in reflection geometry. We focused linearly polarized 60-fs mid-IR pulses at 3.5 $\mu$m) onto the (100) surface of GaAs with the linear polarization directed along the [001] axis. We prepared samples polished on both sides with thicknesses of 45 $\mu$m and 170 $\mu$m. We detected the HHG spectra without spatial filtering. The oscillatory behavior in the field-intensity dependence of the 5th-harmonic integrated intensity was much clearer when the reflection geometry was used but it was faded when the transmission geometry was used despite the initial increase with the $E^{10}$ perturbative scaling. For the 170-$\mu$m-thick sample, the oscillatory behavior completely disappeared. These differences are considered to be caused by the self-focusing effect\cite{Vampa2018_S} and the nonlinear absorption\cite{Xia2018_S} of the mid-IR pulses during propagation inside the sample.

The propagation length $L_{c}$ of self-focusing Gaussian beams until collapse in GaAs was evaluated to be 180 $\mu$m from the formula\cite{Couairon2007_S} $L_{c}\simeq 0.367 L_{\mathrm{DF}}/\sqrt{P_{\mathrm{in}} / P_{\mathrm{cr}} }$, where the Rayleigh length $L_{\mathrm{DF}}$ is 10 mm, the peak power of the mid-IR pulses $P_{\mathrm{in}}$ is 20 MW, and the critical power $P_{\mathrm{cr}}$ for self-focusing is 0.05 MW. The calculation of these parameters was based on our experimental conditions with an $f$-number of $f$/32 and a peak field strength of 3 MV/cm on the sample surface and by using the nonlinear refractive index $n_2=10^{-13}$ cm$^2$/W of GaAs\cite{Hurlbut2007_S}. Moreover, peak field strengths higher than 3 MV/cm caused a drastic decrease in the transmittance of the mid-IR pulses (Fig. \ref{figS6}(b) (gray line)), because of the nonlinear absorption in GaAs. Therefore, we consider that the self-focusing effect of the mid-IR pulses during the propagation {length} ($\sim$100 $\mu$m) initially enhances the HHG intensity in the perturbative regime, and at peak field strengths higher than 3 MV/cm corresponding to the saturation regime, the nonlinear absorption coupled with the self-focusing effect causes the oscillatory behavior to fade.

\section{Experimental results for wavelength, pulse width, and CEP dependence of harmonic spectral shape}
\begin{figure}[t]
\includegraphics[width=140mm]{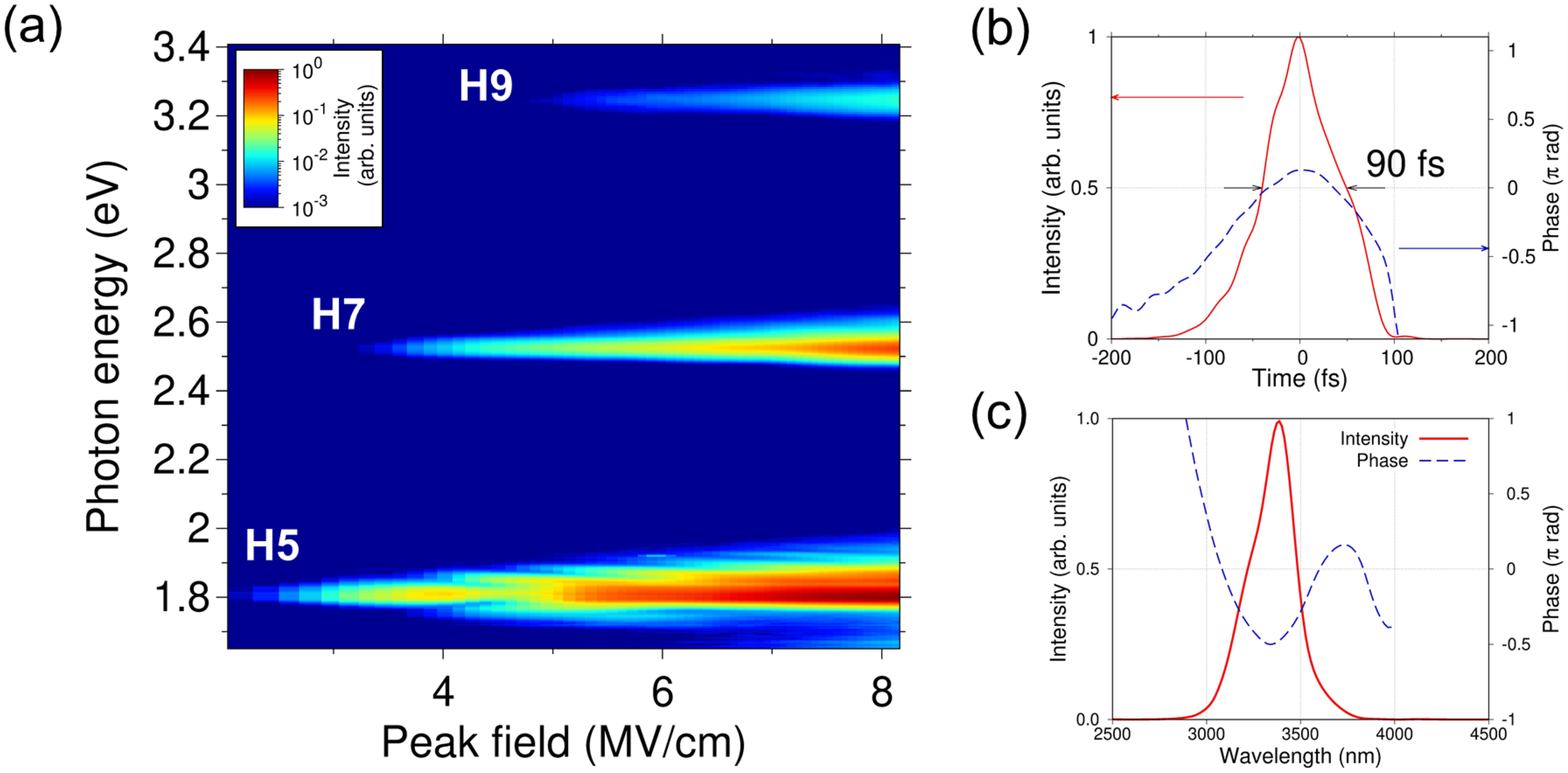}
\caption{ (a) Field-intensity dependences of the 5th, 7th, and 9th harmonic spectra. (b) Temporal profile and (c) spectral shape of the multi-cycle MIR pulses at 3.4 $\mu$m. }
\label{figS2}
\end{figure}

In this section, we show the experimental results of HHG driven by mid-IR pulses for different wavelengths and pulse widths from those at 3.65 $\mu$m and 80 fs in the main text. In addition, we show the carrier-envelope phase (CEP) dependence of the HHG spectrum measured when a few-cycle mid-IR source was used.

Figure \ref{figS2}(a) shows the results when we tuned the center wavelength at 3.4 $\mu$m (0.36 eV) by changing the phase matching condition in optical parametric amplification as shown in Fig. \ref{figS2}(b)(c). The temporal profile was characterized by second harmonic generation frequency-resolved optical gating (SHG-FROG) with a duration of 90 fs (8-optical-cycle). As shown in Fig. \ref{figS2}(a), non-Gaussian spectral shapes appeared in the 5th harmonic spectra ($\sim$1.8 eV); the shapes depended on the laser intensity, but differed from those in the main text, i.e., those of the 5th harmonic ($\sim$1.7 eV) driven by the mid-IR pulses at 3.65 $\mu$m.

\begin{figure}[t]
\includegraphics[width=140mm]{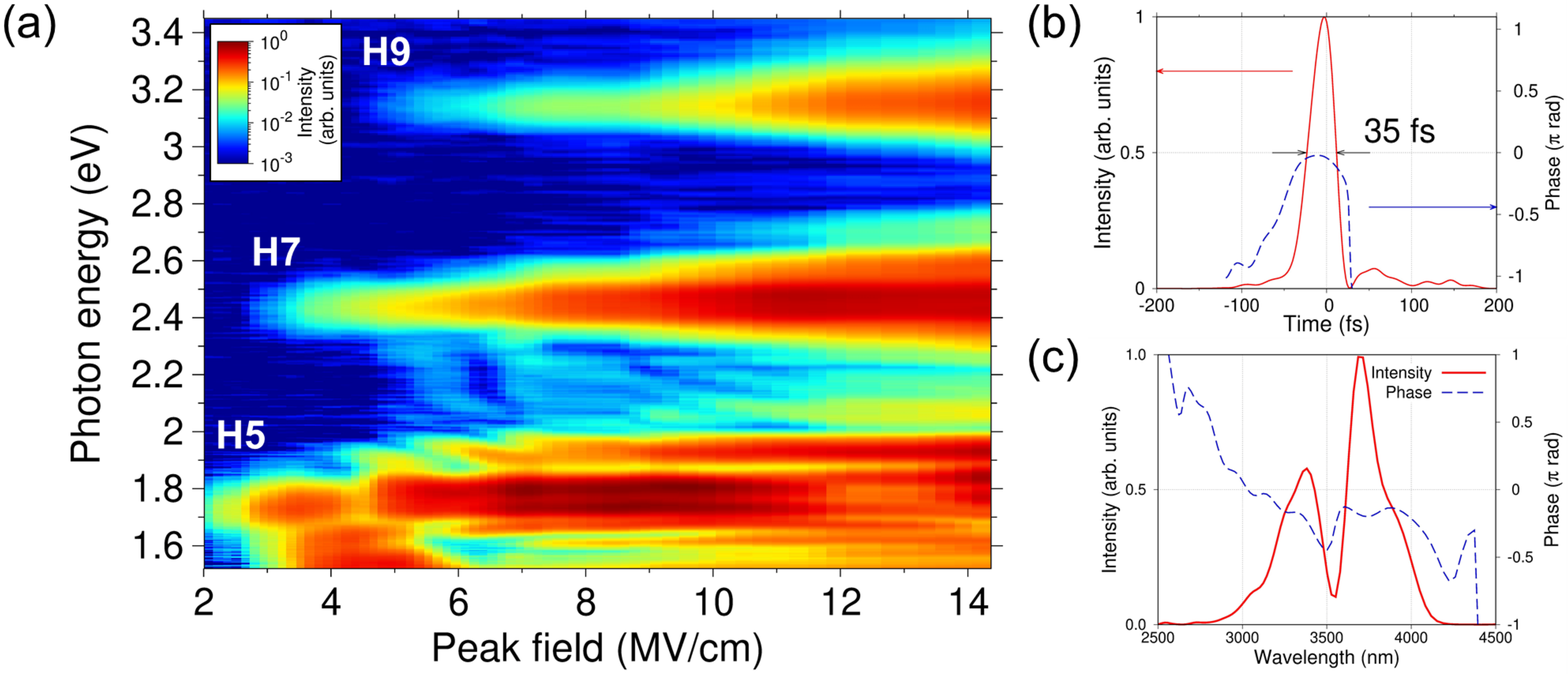}
\caption{ (a) Field-intensity dependences of the 5th, 7th, and 9th harmonic spectra. (b) Temporal profile and (c) spectral shape of the 3-optical-cycle MIR pulses. }
\label{figS3}
\end{figure}

\begin{figure}[t]
\includegraphics[width=100mm]{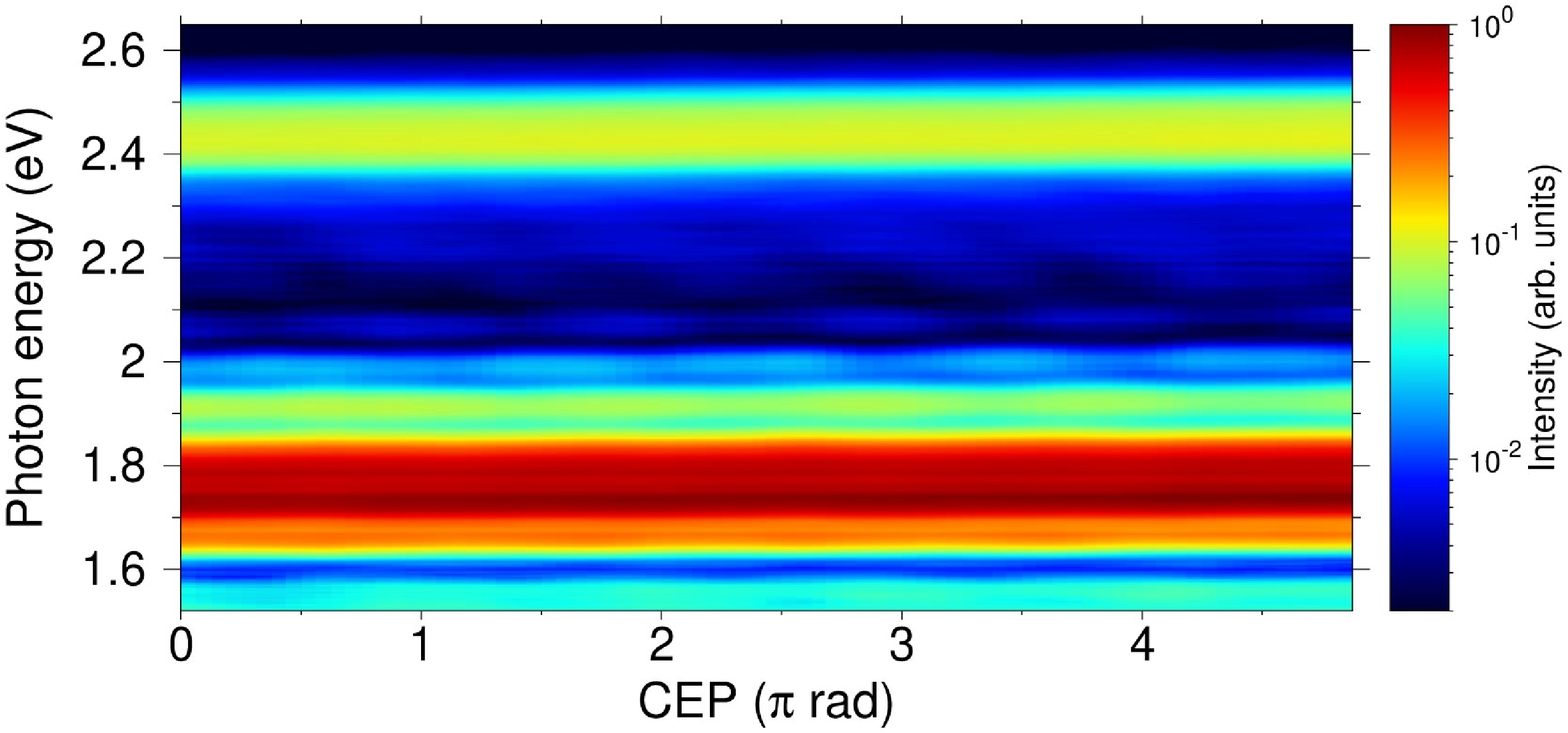}
\caption{ CEP dependence in the 5th- and 7th-harmonic spectral ranges, measured using 3-optical-cycle mid-IR pulses at a peak field strength of 6.5 MV/cm }
\label{figS4}
\end{figure}

Next, we show results for 35-fs (3-optical-cycle) mid-IR pulses at a center wavelength of 3.5 $\mu$m (Fig. \ref{figS3}(a)). To generate these pulses, we used spectral broadening in Si and YAG plates \cite{LuOL2018_S} (see Fig. \ref{figS3}(b)(c)). To avoid undesired interference between the HHG spectrum and the supercontinuum spectrum of the fundamental pulses, we controlled the amount of spectral broadening within the IR spectral range. The damage threshold of the GaAs sample was estimated to be up to 1.4 TW/cm$^2$ in vacuum (15 MV/cm inside the sample) for this few-cycle source. As shown in Fig. \ref{figS3}(a), the 5th harmonic spectrum had more than two peaks (different from Fig. \ref{figS2}(a) and Fig. 1(c) in the main text). Figure \ref{figS4} shows the CEP dependence of the peaks. The CEP of the mid-IR pulses was changed by a CaF$_2$ wedge pair with a stability of $\sim$300 mrad \cite{LuOL2018_S}. The peaks from 1.6 to 1.9 eV were not sensitive to the CEP. The $\pi$-period oscillation in the range 1.9 - 2.3 eV seemed to be caused by interference between the 5th and 7th harmonic spectral edges. Consequently, we conclude that the CEP of the mid-IR pulses had negligible effect on the harmonic spectral shapes, while the wavelength and pulse width had large effects.

The dependence on the drive-laser waveform indicates that the shape of the harmonic spectrum is sensitive to the electronic states in GaAs created by the multiphoton and tunnel ionization process. In particular, the 5th harmonic spectra are considered to be stressed by five-photon resonance with an energy gap (1.74 eV) between the conduction and split-off bands of GaAs. 

\bibliography{sup}{}{}

\begin{thebibliography}{99}
\bibitem{Corkum1993} 
P. B. Corkum, Phys. Rev. Lett.71, 1994 (1993).
\bibitem{Protopapas1997} 
M. Protopapas, C. H. Keitel, and P. L.  Knight, Rep. Prog. Phys. {\bf 60}, 389 (1997).
\bibitem{Brabec2000} 
T. Brabec and F. Krausz, Rev. Mod. Phys. {\bf 72}, 545 (2000).
\bibitem{Agostini2004} P. Agostini and L. F. DiMauro, Rep. Prog. Phys. {\bf 67}, 813 (2004).
\bibitem{Curkum2007} 
P. B. Corkum and F. Krausz, Nature Phys.3, 381 (2007).
\bibitem{Krausz2009} 
F. Krausz and M. Ivanov, Rev. Mod. Phys. {\bf 81}, 163 (2009).
\bibitem{Ghimire2011} 
S. Ghimire, A. D. DiChiara, E. Sistrunk, P. Agostini, L. F. DiMauro, and D. A. Reis, Nature  Phys. {\bf 7}, 138 (2011).
\bibitem{Schubert2014} 
O. Schubert, M. Hohenleutner, F. Langer, B. Urbanek, C. Lange, U. Huttner, D. Golde, T. Meier, M. Kira, S. W.Koch, and R. Huber, Nature Photon. {\bf 8}, 119 (2014).
\bibitem{Luu2015} 
T. T. Luu, M. Garg, S. Y. Kruchinin, A. Moulet, M. T.Hassan,  and E. Goulielmakis, Nature {\bf 521}, 498 (2015).
\bibitem{Hohenleutner2015} 
M. Hohenleutner, F. Langer, O. Schubert, M. Knorr, U. Huttner, S. Koch, M. Kira, and R. Huber, Nature {\bf 523}, 572 (2015).
\bibitem{Vampa2015} 
G. Vampa, T. J. Hammond, N. Thir{\'e}, B. E. Schmidt, F. L{\'e}gar{\'e}, C. R. McDonald, T. Brabec, P. B. Corkum, Nature {\bf 522}, 462 (2015).
\bibitem{Ghimire2019} 
S. Ghimire and D. A. Reis, Nature Phys. {\bf 15}, 10 (2019).
\bibitem{Ndabashimiye2016} 
G. Ndabashimiye, S. Ghimire, M. Wu, D. A. Browne, K. J. Schafer, M. B. Gaarde, and D. A. Reis, Nature {\bf 534}, 520 (2016).
\bibitem{Lanin2017} 
A. A. Lanin, E. A. Stepanov, A. B. Fedotov, and A. M.Zheltikov, Optica {\bf 4}, 516 (2017)
\bibitem{Liu2017} 
H. Liu, Y. Li, Y. S. You, S. Ghimire, T. F. Heinz, and D. A. Reis, Nature Phys. {\bf 13}, 262 (2017).
\bibitem{You2017} 
Y. S. You, D. A. Reis, and S. Ghimire, Nature Phys. {\bf 13}, 345 (2017).
\bibitem{Yoshikawa2017} 
N. Yoshikawa, T. Tamaya,  and K. Tanaka, Science {\bf 356}, 736 (2017).
\bibitem{Kim2017} 
H. Kim, S. Han, Y. W. Kim, S. Kim, and S.-W. Kim, ACS Photonics {\bf 4}, 1627 (2017).
\bibitem{Jiang2018} 
S. Jiang, J. Chen, H. Wei, C. Yu, R. Lu, and C. D. Lin, Phys. Rev. Lett. {\bf 120}, 253201 (2018).
\bibitem{Langer2018} 
F. Langer, C. Schmid, S. Schlauderer, M. Gmitra, J. Fabian, P. Nagler, C. Sch\"ullerr, T. Korn, P. Hawkins, J. Steiner, U. Huttner, S. Koch, M. Kira, and R. Huber, Nature {\bf 557}, 76 (2018).
\bibitem{VampaOE2018} 
G. Vampa and Y. S. You and H. Liu and S. Ghimire and D. A. Reis, Opt. Express {\bf 26}, 12210 (2018).
\bibitem{Silva2018} 
R. E. F. Silva, I. V. Blinov, A. N. Rubtsov, O. Smirnova,and M. Ivanov, Nature Photon. {\bf 12}, 266 (2018).
\bibitem{Hirori2019} 
H. Hirori, P. Xia, Y. Shinohara, T. Otobe, Y. Sanari, H. Tahara, N. Ishii, J. Itatani, K. L. Ishikawa, T. Aharen, M. Ozaki, A. Wakamiya, and Y. Kanemitsu, APL Mater. {\bf 7}, 041107 (2019).
\bibitem{Cheng2020} 
B. Cheng, N. Kanda, T. N. Ikeda, T. Matsuda, P. Xia, T. Schumann, S. Stemmer, J. Itatani, N. P. Armitage, and R. Matsunaga, Phys. Rev. Lett. {\bf 124}, 117402 (2020).
\bibitem{Yariv1984} 
A. Yariv and P. Yeh,Optical Waves in Crystals(1984).
\bibitem{Shen1984} 
Y. R. Shen, The Principles of Nonlinear Optics (1984).
\bibitem{Boyd2008} 
R. W. Boyd, {\it Nonlinear Optics}, 3rd ed. (2008).
\bibitem{Gohle2005} 
C. Gohle, T. Udem, M. Herrmann, J. Rauschenberger, R. Holzwarth, H. A. Schuessler, F. Krausz, and T. W. H\"ansch, Nature {\bf 436}, 234 (2005).
\bibitem{Zair2008} 
A. Za\"ir, M. Holler, A. Guandalini, F. Schapper, J. Biegert, L. Gallmann, U. Keller, A. S. Wyatt, A. Mon-mayrant, I. A. Walmsley, E. Cormier, T. Auguste, J. P. Caumes,  and P. Sali\`eres, Phys. Rev. Lett. {\bf 100}, 143902 (2008).
\bibitem{Yost2009} 
D. C. Yost, T. R. Schibli, J. Ye, J. L. Tate, J. Hostetter, M. B. Gaarde, and K. J. Schafer, Nature Phys. {\bf 5}, 815 (2009).
\bibitem{Cingoz2012} 
A. Cing\"oz, D. C. Yost, T. K. Allison, A. Ruehl, M. E. Fermann, I. Hartl, and J. Ye, Nature {\bf 482}, 68 (2012).
\bibitem{Balcou1999} 
P. Balcou, A. S. Dederichs, M. B. Gaarde, and A. L’Huillier, J. Phys. B: At. Mol. Opt. Phys. {\bf 32}, 2973 (1999).
\bibitem{Popruzhenko2002} 
S. V. Popruzhenko, P. A. Korneev, S. P. Goreslavski, and W. Becker, Phys. Rev. Lett. {\bf 89}, 023001 (2002).
\bibitem{Kopold2002} 
R. Kopold, W. Becker, M. Kleber, and G. G. Paulus, J.Phys. B: At. Mol. Opt. Phys. {\bf 35}, 217 (2002).
\bibitem{Ishikawa2009} 
K. L. Ishikawa, K. Schiessl, E. Persson, and J. Durgd\"orfer, Phys. Rev. A {\bf 79}, 033411 (2009).
\bibitem{Chin2000} 
A. H. Chin, J. M. Bakker, and J. Kono, Phys. Rev. Lett. {\bf 85}, 3293 (2000).
\bibitem{Eyres2001} 
L. A. Eyres, P. J. Tourreau, T. J. Pinguet, C. B. Ebert, J. S. Harris, M. M. Fejer, L. Becouarn, B. Gerard, and E. Lallier, Appl. Phys. Lett. {\bf 79}, 904 (2001).
\bibitem{Mucke2001} 
O. D. M\"ucke, T. Tritschler, M. Wegener, U. Morgner, and F. X. K\"artnerr, Phys. Rev. Lett. {\bf 87}, 057401 (2001).
\bibitem{Hirori2011} 
H. Hirori, K. Shinokita, M. Shirai, S. Tani, Y. Kadoya, and K. Tanaka, Nature Commun.2, 594 (2011).
\bibitem{Zaks2012} 
B. Zaks, H. Banks, and M. Sherwin, Appl. Phys. Lett. {\bf 102}, 012104 (2012).
\bibitem{Fan2013} 
K. Fan, H. Y. Hwang, M. Liu, A. C. Strikwerda, A. Sternbach, J. Zhang, X. Zhao, X. Zhang, K. A. Nelson, and R. D. Averitt, Phys. Rev. Lett. {\bf 110}, 217404 (2013).
\bibitem{Wismer2016} 
M. S. Wismer, S. Y. Kruchinin, M. Ciappina, M. I. Stockman,  and V. S. Yakovlev, Phys. Rev. Lett. {\bf 116}, 197401 (2016).
\bibitem{Liu2016}
S. Liu, M. B. Sinclair, S. Saravi, G. A. Keeler, Y. Yang, J. Reno, G. M. Peake, F. Setzpfandt, I. Staude, T. Pertsch, and I. Brener, Nano Lett. {\bf 16}, 5426 (2016).
\bibitem{Schmidt2018} 
C. Schmidt, J. Btesihler, A.-C. Heinrich, J. Allerbeck, R. Podzimski, D. Berghoff, T. Meier, W. G. Schmidt, C. Reichl, W. Wegscheider, D. Brida, and A. Leitenstorfer, Nature Commun. {\bf 9}, 2890 (2018).
\bibitem{Schlaepfer2018} 
F. Schlaepfer, M. Lucchini, S. A. Sato, M. Volkov, L. Kasmi, N. Hartmann, A. Rubio, L. Gallmann, and U. Keller, Nature Phys. {\bf 14}, 560 (2018).
\bibitem{Ghalgaoui2018} 
A. Ghalgaoui, K. Reimann, M. Woerner, T. Elsaesser, C. Flytzanis, and K. Biermann, Phys. Rev. Lett. {\bf 121}, 266602 (2018).
\bibitem{Xia2018} 
P. Xia, C. Kim, F. Lu, T. Kanai, H. Akiyama, J. Itatani, and N. Ishii, Opt. Express {\bf 26}, 29393 (2018).
\bibitem{Lu2019Nat} 
J. Lu,  E. F. Cunningham,  Y. S. You,  D. A. Reis,   andS. Ghimire, Nature Photon.13, 96 (2019).
\bibitem{Luttinger1955} 
J. M. Luttinger and W. Kohn, Phys. Rev. {\bf 97}, 869 (1955).
\bibitem{Chuang1991} 
S. L. Chuang, Phys. Rev. B43, 9649 (1991).
\bibitem{Ahn1995} 
D. Ahn, S. J. Yoon, S. L. Chuang, and C.-S. Chang, J. Appl. Phys. {\bf 78}, 2489 (1995).
\bibitem{Pfeffer1996} 
P. Pfeffer and W. Zawadzki, Phys. Rev. B {\bf 53},  12813 (1996).
\bibitem{Pryor1998} 
C. Pryor, Phys. Rev. B {\bf 57}, 7190 (1998)
\bibitem{Dargys2002} 
A. Dargys, Phys. Rev. B66, 165216 (2002).
\bibitem{Tomic2006} 
S. Tomi\'c, A. G. Sunderland, and I. J. Bush, J. Mater. Chem. {\bf 16}, 1963 (2006).
\bibitem{Luque2015} 
A. Luque, A. Panchak, A. Mellor, A. Vlasov, A. Mart\'i,and V. Andreev, Sol. Energy Mater. Sol. Cells {\bf 141}, 39 (2015).
\bibitem{Bastos2016} 
C. M. O. Bastos, F. P. Sabino, P. E. F. Junior, T. Campos, J. L. F. D. Silva,  and G. M. Sipahi, Semicond. Sci. Technol. {\bf 31}, 105002 (2016).
\bibitem{Sytnyk2018} 
D. Sytnyk and R. Melnik, arXiv:1808.06988 (2018).
\bibitem{Tamaya2016} 
T. Tamaya, A. Ishikawa, T. Ogawa, and K. Tanaka, Phys. Rev. Lett. {\bf 116}, 016601 (2016).
\bibitem{Tamaya2016PRBR} 
T. Tamaya, A. Ishikawa, T. Ogawa, and K. Tanaka, Phys. Rev. B {\bf 94}, 241107(R) (2016).
\bibitem{Tamaya2019} 
T. Tamaya and T. Kato, Phys. Rev. B {\bf 100}, 081203(R) (2019).
\bibitem{Lu2018} 
F. Lu,  P. Xia,  Y. Matsumoto,  T. Kanai,  N. Ishii,   andJ. Itatani, Opt. Lett.43, 2720 (2018).
\bibitem{Haug2009} 
H. Haug and S. W. Koch, {\it Quantum Theory of the Optical and Electronic Properties of Semiconductors}, 5th ed. (2009).
\bibitem{Bastard1990} 
G. Basterad, {\it Wave Mechanics Applied to Semiconductor Heterostructures} (1990).
\bibitem{Kane1957} 
E. O. Kane, Journal of Physics and Chemistry of Solids1, 249 (1957).
\bibitem{Dunlap1986} 
D. H. Dunlap and V. M. Kenkre, Phys. Rev. B {\bf 34}, 3625 (1986).
\bibitem{Grossmann1991} 
F. Grossmann,  T. Dittrich, P. Jung, and P. H\"anggi, Phys. Rev. Lett. {\bf 67}, 516 (1991).
\bibitem{Oka2005} 
T. Oka, R. Arita,   and H. Aoki, Physica B Condens. Matter {\bf 359-361}, 759 (2005).
\bibitem{Lignier2007} 
H. Lignier, C. Sias, D. Ciampini, Y. Singh, A. Zenesini, O. Morsch, and E. Arimondo, Phys. Rev. Lett. {\bf 99}, 220403 (2007).
\end{thebibliography}

\section{References}
\begin{thebibliography}{99}
\bibitem{Tamaya2016PRL_S} 
T. Tamaya, A. Ishikawa, T. Ogawa, and K. Tanaka, Phys. Rev. Lett. {\bf{116}}, 016601 (2016).
\bibitem{Tamaya2016PRBR_S} 
T. Tamaya, A. Ishikawa, T. Ogawa, and K. Tanaka, Phys. Rev. B {\bf{94}}, 241107(R) (2016).
\bibitem{Tamaya2017Science_S} 
N. Yoshikawa, T. Tamaya, and K. Tanaka, Science {\bf{356}}, 736 (2017).
\bibitem{Tamaya2019PRBR_S} 
T. Tamaya and T. Kato, Phys. Rev. B {\bf{100}}, 081203(R) (2019).
\bibitem{Luttinger_S} 
J. M. Luttiner and W. Kohn, Phys. Rev {\bf{97}}, 4 (1955).
\bibitem{Chuang_S} 
S. L. Chuang, Phys. Rev. B {\bf{43}} 12 (1991).
\bibitem{Ahn_S} 
D. Ahn, S. J. Yoon, S. L. Chuang, and C.-S. Chang, J. Appl. Phys. {\bf{78}}, 2489 (1995).
\bibitem{Dargys_S} 
A. Dargys, Phys. Rev. B {\bf{66}} 165216 (2002).
\bibitem{Pryor_S} 
C. Pryor, Phys. Rev. B {\bf{57}} 7190 (1997).
\bibitem{Luque_S} 
A. Luque, A. Panchak, A. Mellor, A. Vlasov, A. Mart\'{i}, and V. Andreev, Sol. Energy Mater. Sol. Cells, {\bf{141}} 39 (2015).
\bibitem{Tomic_S} 
S. Tomi\`c, A. G. Sunderland, and I. J. Bush, J. Mater. Chem. {\bf{16}},1963 (2006).
\bibitem{Sytnyk_S} 
D. Sytnyk and R. Melnik, arXiv:1808.06988.
\bibitem{Bastos_S} 
Carlos M O Bastos, Fernando P Sabino, Paulo E Faria Junior, Tiago Campos, Juarez L F Da Silva, and Guilherme M Sipahi, Semicond. Sci. Technol. {\bf{31}} 105002 (2016).
\bibitem{Pfeffer_S} 
P. Pfeffer and W. Zawadzki, Phys. Rev. B {\bf{53}} 12813 (1996). 
\bibitem{Floss2018_S} 
I. Floss, C. Lemell, G. Wachter, V. Smejkal, S. A. Sato, X.-M. Tong, K. Yabana, and J. Burgd\"orfer, Phys. Rev. A {\bf{97}}, 011401 (2018).
\bibitem{Abadie2018_S} 
C. Q. Abadie, M. Wu, and M. B. Gaarde, Opt. Lett. {\bf{43}}, 5339 (2018).
\bibitem{Zair2008_S} 
A. Za\"{\i}r, M. Holler, A. Guandalini, F. Schapper, J. Biegert, L. Gallmann, U. Keller, A. S. Wyatt, A. Mon-mayrant, I. A. Walmsley, E. Cormier, T. Auguste, J. P. Caumes, and P. Sali\`eres, Phys. Rev. Lett. {\bf{100}}, 143902 (2008).
\bibitem{Toma1999_S} 
E. S. Toma, P. Antoine, A. de Bohan, and H. G. Muller, J. Phys. B: At. Mol. Opt. Phys. {\bf{32}}, 5843 (1999).
\bibitem{Vampa2018_S} 
G. Vampa and Y. S. You and H. Liu and S. Ghimire and D. A. Reis, Opt. Express {\bf{26}}, 12210 (2018).
\bibitem{Xia2018_S} 
P. Xia, C. Kim, F. Lu, T. Kanai, H. Akiyama, J. Itatani, and N. Ishii, Opt. Express {\bf{26}}, 29393 (2018).
\bibitem{Couairon2007_S} 
A. Couairon and A. Mysyrowicz, Physics Reports {\bf{441}}, 53 (2007).
\bibitem{Hurlbut2007_S} 
W. C. Hurlbut and Y.-S. Lee, Opt. Lett. {\bf{32}}, 668 (2007).
\bibitem{LuOL2018_S} 
F. Lu,  P. Xia,  Y. Matsumoto,  T. Kanai,  N. Ishii, and J. Itatani, Opt. Lett. {\bf{43}}, 2720 (2018).
\end{thebibliography}

\end{document}